\documentclass[
11pt,
a4paper,
oneside, 
headinclude,footinclude, 
BCOR5mm,
]{article}
\usepackage{tgtermes}
\usepackage[utf8]{inputenc}
\usepackage{amsmath}
\usepackage{subcaption}
\usepackage{caption}
\usepackage{longtable}
\usepackage{amsthm}
\usepackage{color}
\usepackage[dvipsnames]{xcolor}
\usepackage{diagbox}
\usepackage{hyperref}
\usepackage{authblk}
\hypersetup{
    colorlinks=true, 
    linkcolor=blue, 
    urlcolor=red, 
    linktoc=all 
}
\usepackage{amssymb}
\usepackage{pagecolor,lipsum}
\usepackage{enumerate}
\usepackage{pdfpages}
\usepackage{xcolor}
\usepackage{tikzsymbols}
\usepackage{blkarray}
\usepackage{enumitem}
\usepackage{graphicx}
\usepackage{setspace}
\usepackage{soul}
\usepackage{enumitem}
\usepackage{soul}
\usepackage{MnSymbol}
\usepackage{mathtools}
\usepackage{bbm}
\usepackage{braket}
\usepackage[framemethod=tikz]{mdframed}
\usepackage{amsmath}
\usepackage[thinc]{esdiff}
\usepackage{array}
\usepackage{booktabs}
\usepackage{algorithm}
\usepackage{algpseudocode}
\usepackage{geometry}
\usepackage{geometry}
 \let\oldpara\paragraph
\renewcommand{\paragraph}[1]{\vspace{-0.5cm} \oldpara{#1}}

\definecolor{mycolor}{rgb}{0.122, 0.435, 0.698}
\newmdenv[topline=false, bottomline=false, rightline=false, innerlinewidth=0.4pt, roundcorner=4pt,linecolor=mycolor,innerleftmargin=6pt,
innerrightmargin=6pt,innertopmargin=1pt,innerbottommargin=6pt]{mybox}

\newmdenv[backgroundcolor=gray!10, topline=false, bottomline=false, rightline=false, innerlinewidth=0.4pt, roundcorner=4pt,linecolor=black,innerleftmargin=6pt,
innerrightmargin=6pt,innertopmargin=3pt,innerbottommargin=6pt]{mybox2}

\newmdenv[backgroundcolor=blue!5, topline=false, bottomline=false, rightline=false, leftline=false, innerlinewidth=0.4pt, roundcorner=4pt,innerleftmargin=10pt,
innerrightmargin=10pt,innertopmargin=10pt,innerbottommargin=10pt]{mybox3}

\usepackage{sectsty}
\definecolor{darkblue}{RGB}{0,0,76}
\title{\textbf{Networked Anti-Coordination Games Meet Graphical Dynamical Systems: Equilibria and Convergence}}

\author { \small
    Zirou Qiu,\textsuperscript{1,2}
    Chen Chen,\textsuperscript{2}
    Madhav V. Marathe,\textsuperscript{1,2}
    S. S. Ravi,\textsuperscript{2,3}\\
    Daniel J. Rosenkrantz,\textsuperscript{2,3}
    Richard E. Stearns,\textsuperscript{2,3}
    Anil Vullikanti\textsuperscript{1,2}
}
\date{}

\affil[1]{\small Computer Science Dept., University of Virginia.}
\affil[2]{\small Biocomplexity Institute and Initiative, University of Virginia.}
\affil[3]{\small Computer Science Dept., University at Albany – SUNY.}

\newcommand{\mc}[1]{\mathcal{#1}}

\newcommand{\ts}[1]{\textsc{#1}}
\newcommand{\eqe}[1]{\textsc{EQE}}
\newcommand{\eqf}[1]{\textsc{EQF}}
\newcommand{\conv}[1]{\textsc{Conv}}
\newcommand{\eqc}[1]{\textsc{EQC}}
\newcommand{\ep}[1]{\mathbb{E}}
\newcommand{\done}[1]{\textcolor{blue}{(\textbf{Done})}}

\newcommand{\pr}[1]{\mathbb{Pr}}

\DeclareMathAlphabet{\mymathbb}{U}{BOONDOX-ds}{m}{n}
\newcommand{\szero}[1]{\mymathbb{0}}
\newcommand{\sone}[1]{\mymathbb{1}}
\newcommand{\z}[1]{\mathbb{Z}}
\newcommand{\nminfpe}[1]{\textsc{NMin-FPE}}
\newcommand{\nmaxfpe}[1]{\textsc{NMax-FPE}}
\newcommand{\mvcG}[1]{G_{\mathcal{M}}}
\newcommand{\nminfpr}[1]{\textsc{NMin-FPR}}
\newcommand{\nmaxfpr}[1]{\textsc{NMax-FPR}}

\newcommand{\zt}[1]{\mathbb{Z}_T}

\newcommand{\sydsG}{G_{\mc{S}}}
\newcommand{\sdsG}{G_{\mc{S'}}}

\newtheorem{theorem}{\textbf{Theorem}}[section]

\newtheorem{lemma}[theorem]{\textbf{Lemma}}
\newtheorem{definition}[theorem]{\textbf{Definition}}

\newtheorem{observation}[theorem]{Observation}

\newenvironment{proofsketch}{%
  \proof}{\endproof}

\newcommand{\snsyds}{\textsf{(SN, IT)-SyDS}}
\newcommand{\sesyds}{\textsf{(SE, IT)-SyDS}}
\newcommand{\snsds}{\textsf{(SN, IT)-SDS}}
\newcommand{\sesds}{\textsf{(SE, IT)-SDS}}
\newcommand{\syds}{\textsf{SyDS}}
\newcommand{\sds}{\textsf{SDS}}

\newcommand{\syacg}{\textsf{SyACG}}
\newcommand{\sqacg}{\textsf{SACG}}
\newcommand{\snsyacg}{\textsf{SN-SyACG}}
\newcommand{\sesyacg}{\textsf{SE-SyACG}}
\newcommand{\snsqacg}{\textsf{SN-SACG}}
\newcommand{\sesqacg}{\textsf{SE-SACG}}
\newcommand{\se}{\textsf{SE}}
\newcommand{\sn}{\textsf{SN}}

\newcommand{\cals}{\mbox{$\mathcal{S}$}}

\newcommand{\withmemory}{\textit{self essential}}

\newcommand{\withoutmemory}{\textit{self non-essential}}

\newcommand{\Ct}[1]{C^{#1}}

\begin{document}

\maketitle

\begin{abstract}
\noindent
Evolutionary anti-coordination games on networks capture real-world strategic situations such as traffic routing and market competition. In such games, agents maximize their utility by choosing actions that differ from their neighbors' actions. Two important problems concerning evolutionary games are the existence of a pure Nash equilibrium (NE) and the convergence time of the dynamics. In this work, we study these two problems for anti-coordination games under {\em sequential} and {\em synchronous} update schemes. 
For each update scheme, we examine two decision modes based on whether an agent considers its own previous action  (\withmemory{}) or not (\withoutmemory{}) in choosing its next action. 
Using a connection to dynamical systems, we show that for both update schemes, finding an NE can be done efficiently under the \withoutmemory{} mode but is computationally intractable under the \withmemory{} mode.
To cope with this hardness, we identify special cases for which an NE can be obtained efficiently.
For convergence time, we show that the best-response dynamics converges in a polynomial number of steps in the synchronous scheme for both modes; for the sequential scheme, the convergence time is polynomial only under the \withoutmemory{} mode. Through experiments, we empirically examine the convergence time and the equilibria for both synthetic and real-world networks.

\smallskip
\noindent
\textbf{Conference version.} The conference version of the paper is accepted at \texttt{\textbf{AAAI-2023}}: \href{https://ojs.aaai.org/index.php/AAAI/article/view/26378}{\textbf{Link}}. 
\end{abstract}

\section{Introduction}
{\em Evolutionary} anti-coordination (AC) games have been widely used to model real-world strategic situations such as product marketing~\cite{linde2014strategies}, balanced traffic routing~\cite{galib2011road}, and social competition~\cite{bramoulle2007anti}. 
In the {\em networked} version of such a game, vertices are agents (players), and edges are interactions between agents. At each time step, agents make new decisions based on the decisions of their neighbors~\cite{young2015evolution}. Specifically, under the \textit{best-response dynamics} of an anti-coordination game with binary actions, each agent maximizes its utility at the current time step by choosing a particular action (i.e., $0$ or $1$) if and only if a sufficient number of its neighbors chose the \textbf{opposite} action at the previous step~\cite{ramazi2016networks}. 
There are two main types of update schemes for evolutionary games, where agents either choose actions \textit{synchronously} or \textit{sequentially}~\cite{adam2012behavior}. 

\par The decision mode where each agent only considers its neighbors' actions in making its decisions is common in the game theory literature~\cite{gibbons1992primer}. Nevertheless, in real-world situations where agents compete for resources, it is natural for an agent to also consider its {\em own} previous action before choosing a new action~\cite{gibbons1992primer,traulsen2010human,lo2006game}.
For example, drivers on a highway can be seen as agents in an evolutionary anti-coordination game, where people switch lanes to avoid traffic. In particular, an agent's choice of lanes in the next time step is influenced by both its current lane and the lanes of neighboring cars. 
Such considerations motivate us to investigate two decision modes: ($i$) \withmemory{} (\textbf{\se{}}), where each agent considers both its previous action and the actions of its neighbors, and $(ii)$ \withoutmemory{} (\textbf{\sn{}}), where each agent only considers the actions of its neighbors\footnote{Our usage of the word ``essential" here is based on the term ``essential variable" used in the context of Boolean (and other) functions to indicate the dependence of a function on a variable~\cite{Salomaa-1963}.}.

Pure Nash equilibria (NE) are a central concept in game theory~\cite{aleksandrov2017pure,Nisan-etal-2007}.
In evolutionary games, another key notion is the time it takes for the best-response dynamics to reach an NE or a limit cycle~\cite{christodoulou2006convergence, jackson2015games}. 
Nevertheless, researchers have given limited attention to efficiently finding an NE for anti-coordination games under the \se{} and \sn{} modes. Further, to our knowledge, whether the best-response dynamics of anti-coordination games has a polynomial convergence time remains open. 
In this work, we close the gap with a systematic study of the following two problems for the synchronous and sequential games under both \se{} and \sn{} modes: $(i)$ \ts{Equilibrium existence/finding (\eqe{}/\eqf{}):} Does the game have an NE, and if so, can we {\em find} one efficiently? $(ii)$ \ts{Convergence (Conv):} Starting from an action profile, how fast does the best-response dynamics converge\footnote{We use {\em convergence} to mean that the dynamics reaches a limit cycle of length at most $2$ (e.g., an NE is a limit cycle of length $1$). For games considered in this work, it is known that the length of each limit cycle is at most $2$, except for \se{} sequential AC games, where there can be limit cycles of exponential length.} to a cycle of length at most $2$?

\par The best-response dynamics of an evolutionary anti-coordination game can be specified using a threshold framework~\cite{adam2012threshold}. This naturally allows us to model such a game as a \textbf{graphical dynamical system}~\cite{barrett2006complexity}. 
In such a system, at each time step, each vertex uses its \textit{local function} to compute its state in the next time step. 
To model anti-coordination games, the domain is Boolean, and the local functions are \textit{inverted-threshold functions} whereby each vertex $u$ is assigned state $1$ for the next step if and only if enough neighboring vertices of $u$ are currently in state $0$.
Graphical dynamical systems are commonly used to model the propagation of contagions and decision-making processes over networks~\cite{barrett2006complexity}. Here, we use dynamical systems with inverted-threshold functions as a theoretical framework to study \eqe{}/\eqf{} and \ts{Conv} problems for evolutionary networked anti-coordination games. 

\smallskip
\noindent
\textbf{Main Contributions}

\smallskip
\begin{itemize}[leftmargin=*,noitemsep,topsep=0pt]
    \item \textbf{Finding an NE.} We demonstrate an interesting contrast in the complexity of \eqe{}/\eqf{} between the \se{} and \sn{} modes for anti-coordination games. In particular, we show that \eqe{}/\eqf{} is \textbf{NP}-hard under the \se{} mode for both synchronous and sequential games, even on bipartite graphs. Further, we show that the corresponding counting problem is \textbf{\#P}-hard. On the other hand, we can find an NE efficiently under the \sn{} mode for synchronous and sequential games. We also identify special cases (e.g., the underlying graph is a DAG, or even-cycle free) of \eqe{}/\eqf{} for the \se{} mode where an NE can be efficiently found. 

    \item \textbf{Convergence.} We show that starting from an arbitrary profile, the best-response dynamics of \textit{synchronous} anti-coordination games under either \se{} or \sn{} mode converge in $O(m)$ time steps, where $m$ is the number of edges in the underlying graph. Further, we establish a similar $O(m)$ bound on the convergence time for \textit{sequential} anti-coordination games under the \sn{} mode. We do not consider the convergence problem for the sequential games under the \se{} mode since such systems can have exponentially long cycles \cite{barrett2003some}. 
    
    \item \textbf{Empirical analysis.} We study the contrast in the empirical convergence time for both modes under different classes of networks. Further, we perform simulations to explore how convergence time changes with network density. We also investigate the number of equilibria for problem instances of reasonable sizes.
\end{itemize}

\smallskip

\begin{center}
 \resizebox{0.6\textwidth}{!}{\begin{tabular}{||l | c | c | c | c||} 
 \hline
 \textbf{Problems} & \snsyacg{} & \sesyacg{} & \snsqacg{} & \sesqacg{}\\ [0.5ex] 
 \hline 
  \eqe{} / \eqf{} & Trivial / \textbf{P} & \textbf{NP}-hard & Trivial / \textbf{P} & \textbf{NP}-hard \\ \hline
  
 \conv{} & $O(m)$ & $O(m)$ & $O(m)$ & NA \\ \hline
 
\end{tabular}}
\captionof{table}{\textbf{Overview of key results}. All results, {\em except} those marked with ``Trivial'' and ``NA'', are established in this paper. \textsf{SyACG} (\textsf{SACG}) denotes \textit{synchronous} (\textit{sequential}) anti-coordination game, and \se{} (\sn{}) stands for the \withmemory{} (\withoutmemory{}) mode. The number of edges is $m$. 
The entry ``Trivial" for \eqe{} denotes that the corresponding game always has a NE~\cite{monderer1996potential}.
``NA'' denotes that the problem is not applicable since there can be exponentially long cycles~\cite{barrett2003some}. For \conv{}, $O(m)$ is the number of steps for the best-response dynamics to reach a limit cycle.}
\label{tab:result}
\end{center}

\section{Related Work} 

\paragraph{The existence of NE.} 
The \withoutmemory{} {\em sequential} anti-coordination games are potential games and always have an NE~\cite{vanelli2020games}. This argument follows from~\cite{monderer1996potential}, which guarantees the existence of NE at the maximum potential. Further, Monderer and Shapley~\cite{monderer1996potential} show that general potential games converge in \textit{finite} time. Note that {\em this result does not imply a polynomial-time algorithm in finding an NE}. Kun et al.~\cite{kun2013anti} study a special form of anti-coordination games where each agent chooses the decision that is the opposite of the majority of neighboring decisions, and show that in such a game, an NE can be found efficiently.
Vanelli et al.~\cite{vanelli2020games} examine synchronous games with both coordinating and anti-coordinating agents. They present several special cases on the threshold distributions for the existence of NE. 

\par Auletta et al.~\cite{auletta2016generalized} define the class of generalized discrete preference games and show that such games always have an NE. They also show that every ordinal potential game with binary actions can be reduced to a discrete preference game. Goles and Martinez~\cite{goles2013neural} prove that for synchronous {\em coordination} games, the length of any limit cycle is at most $2$. Many researchers have studied the existence of NE in other games of different forms (e.g., \cite{simon2017constrained,conitzer2002complexity}).

\paragraph{Limiting behavior.} Adam et al.~\cite{adam2012threshold} show that the length of a limit cycle in a synchronous anti-coordination game is at most $2$. However, {\em they did not bound the convergence time to reach a limit cycle.} Ramazi, Riehl, and Cao~\cite{ramazi2016networks} investigate the convergence time for asynchronous \withoutmemory{} anti-coordination games; they establish that an NE will be reached in \textit{finite} time. We note that the asynchronous dynamic they consider is different from the sequential dynamic studied in our work, and their convergence result does not imply ours.

\par Goles and Martinez~\cite{goles2013neural} establish that for {\em coordination games}, the dynamics converges in a polynomial number of steps. Barrett et al.~(\cite{barrett2003some}) study phase space properties of sequential dynamical systems (which includes modeling sequential AC games); the results imply that the length of a limit cycle in a \withmemory{} sequential anti-coordination game can be exponential in the number of agents. The convergence of the best-response dynamics for games of other types has also been studied (e.g., \cite{ieong2005fast,awerbuch2008fast, jackson2015games}.

\paragraph{Minority games.} Anti-coordination games are closely related to minority games~\cite{chmura2006successful,arthur1994inductive}. Many versions of minority games have been studied. For example, Challet and Marsili~\cite{challet2000relevance} study the dynamics of minority games where agents make decisions based on the action profile history of the population. 
Shang et al.~\cite{manuca2000structure} examine the action distribution of minority games over different network topologies. Several other forms of minority games are studied in~\cite{li2000evolution}.

\par A  more thorough discussion of related work is given in the Appendix. To our knowledge, the complexity of \eqf{} for \textsf{SACG}/\textsf{SyACG} has not been established; nor has a polynomial bound on the convergence time of these games been established. We resolve these problems in this work.

\section{Preliminaries and Problem Definition}
A networked game operates on a graph $G$ where vertices are agents and edges represent interactions. At each discrete time step, an agent chooses a \textit{binary} action from the set $\{0, 1\}$ based on neighbors' actions, and receives a payoff. Under the \textit{best-response dynamics}, each agent chooses an action that yields a maximum payoff. 
For the best-response dynamics of an \textbf{anti-coordination} (AC) \textbf{game}, each agent $v$ has a nonnegative threshold $\tau_1(v)$, and $v$ chooses action $1$ if and only if the number of neighbors (plus possibly $v$ itself) that chose action $0$ is at least $\tau_1(v)$~\cite{ramazi2016networks}. We now define the problems of interest.

\smallskip

\begin{mybox2}
\begin{definition}[\textbf{Equilibrium existence/finding}] 
   Given an anti-coordination game, does it have an NE, and if so, can we find it efficiently?
\end{definition} 
\end{mybox2}

\begin{mybox2}
\begin{definition}[\textbf{Convergence}]
    Given an anti-coordination game and an initial action profile, how fast does the best-response dynamics converge to a limit cycle?
\end{definition}
\end{mybox2}

We examine two decision modes, based on whether or not each agent considers its own previous action in making a new decision.
Under the \withoutmemory{} (\textbf{\sn{}}) mode, each agent only considers neighbors' actions, whereas, under the \withmemory{} (\textbf{\se{}}) mode, each agent considers both its own previous action and its neighbors' actions. We further consider two types of update schemes: $(i)$ \textit{synchronous} (\textbf{\textsf{SyACG}}): agents choose actions simultaneously; $(ii)$ \textit{sequential} (\textbf{\textsf{SACG}}): at each time step, agents update their actions in a predefined order. Overall, we examine four classes of AC games based on update scheme (i.e., \textit{synchronous} or \textit{sequential}) and decision mode (i.e., \sn{} or \se{}).

\paragraph{Graphical dynamical systems.} We use \textit{dynamical systems} as a mathematical framework to study anti-coordination games. We follow the notation used in \cite{barrett2006complexity,rosenkrantz2020synchronous}. A \textit{synchronous} dynamical system (\textbf{\syds{}}) over the Boolean state domain $\mathbb{B} = \{0, 1\}$ is a pair $\mc{S} = (\sydsG{}, \mc{F})$; $\sydsG{} = (V,E)$ is the underlying graph, with $n = |V|$ and $m = |E|$. 
We assume $\sydsG{}$ is connected and undirected, unless specified otherwise. The set $\mc{F} = \{f_1, ..., f_n\}$ consists of functions with $f_i$ being the \textit{local function} of vertex $v_i \in V, 1 \leq i \leq n$. The output of $f_i$ gives the state value of $v_i$.
In a \syds{}, vertices update states \textit{simultaneously} at each time step. 

\par A \textit{sequential} dynamical system (\textbf{\sds{}}) $\mc{S}'$ is a tuple $(\sdsG{}, \mc{F}',\Pi)$ where $\sdsG{}$ and $\mc{F}'$ are defined the same as for the above synchronous systems~\cite{mortveit2007introduction}. In addition, $\Pi$ is a permutation of $V$ that determines the sequential order in which vertices update their states at each time step. 
Specifically, each time step of $\mc{S}'$ consists of $n$ {\em substeps}, in each of which one vertex updates its state using the current vertex states in $\mc{S}'$.

\vspace{-0.5cm}
\paragraph{Update rules.} To model anti-coordination dynamics, we consider \textit{inverted-threshold} local functions for the state-update of vertices. Formally, each vertex $v_i$ has a fixed integer threshold $\tau_1(v_i) \geq 0$. Starting from an initial configuration, all vertices update states at each time step using their local functions, and the next state of $v_i$ is $1$ iff the number of neighbors (plus possibly $v_i$ itself) with state-$0$ at the previous time step (for a  \textit{synchronous} scheme) or at the current time step (for a  \textit{sequential} scheme) 
is at least $\tau_1(v_i)$. 

\begin{figure}[!h]
\small
  \centering
\includegraphics[width=0.65\textwidth]{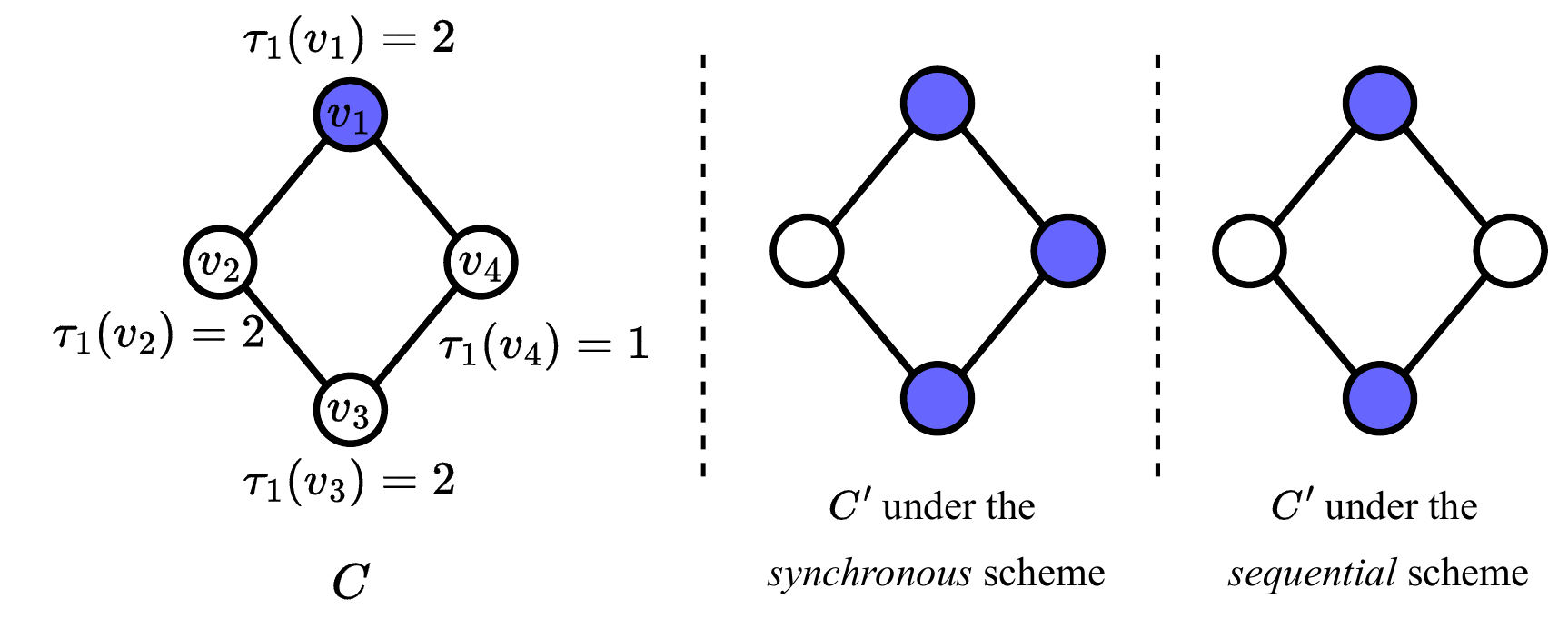}
    \caption{Example dynamics of \sds{} and \syds{} under the \sn{} mode. Specifically, $C$ is an initial configuration, and $C'$ is its successor under either the \textit{synchronous} or the \textit{sequential} (with vertex update order $(v_1, v_2, v_3, v_4)$) mode. State-1 vertices are highlighted in blue.}
    \label{fig:example_syds}
\end{figure}

Lastly, analogous to anti-coordination games, we consider the same two decision modes (i.e., \sn{} and \se{}) for dynamical systems. Based on the schemes (i.e., \syds{} or \sds{}) and decision modes, we have four classes of systems that map to the four classes of anti-coordination games described previously. We use the notation \textsf{(SN/SE, IT)-SDS/SyDS} to denote different classes of systems (\textsf{IT} stands for inverted-threshold). An example of \sn{} mode is shown in Figure~\ref{fig:example_syds}.


\vspace{-0.5cm}
\paragraph{Limit cycles.} A \textit{configuration} $C$ of a system $\mc{S}$ is a vector $C =$ $(C(v_1), ..., C(v_n))$ where $C(v_i) \in \mathbb{B}$ is the state of vertex $v_i$ under $C$. The dynamics of $\mc{S}$ from an initial configuration $C$ can be represented by a time-ordered sequence of configurations. A configuration $C'$ is the \textit{successor} of $C$ if $\mc{S}$ evolves from $C$ to $C'$ in one step. Let $C'$ be the successor of $C$, and $C''$ be the successor of $C'$. If $C = C'$, that is, no vertices undergo state changes from $C$, then $C$ is a \textbf{fixed point} of $\mc{S}$; if $C \neq C'$, but $C = C''$, then $C \rightleftharpoons  C'$ forms a \textbf{2-cycle} of $\mc{S}$.

The dynamics of an AC game is captured by the dynamics of the underlying dynamical system $\mc{S}$. Specifically, an agent $v$'s action at time $t$ corresponds to $v$'s state at time $t$ in $\mc{S}$, and $v$'s decision threshold is described by $\tau_1(v)$. Moreover, the evolution of the action profile for the game coincides with the transition of configurations under $\mc{S}$.

\begin{mybox2}
\begin{observation}
A fixed point and a limit cycle of $\mc{S}$ correspond respectively to an NE and a limit cycle of the action profile for the underlying anti-coordination game.
\end{observation}
\end{mybox2}

\noindent
The convergence time of $\mc{S}$ precisely characterizes the convergence time of the corresponding game. Thus, \textit{proofs of our results for anti-coordination games are given in the context of dynamical systems.}

 \section{Equilibrium Existence and Finding}
 For \withmemory{} (\se{}) anti-coordination games, we establish that \eqe{} (and therefore \eqf{}) is \textbf{NP}-hard, even on bipartite graphs. Further, the corresponding counting problem is \textbf{\#P}-hard. In contrast, for \withoutmemory{} (\sn{}) anti-coordination games, we can find an NE in polynomial time. We remark that the simple difference between the two modes (i.e., whether each agent considers its own state or not) yields a major difference in the complexity of finding an NE. We further discuss the reasons for this contrast in a later section. Lastly, to cope with the hardness under the \se{} mode, we identify special classes where an NE (if it exists) can be found efficiently. 

\par We first observe that if a \syds{} $\mc{S}$ and an \sds{} $\mc{S}'$ have the same underlying graph and the same local functions, then they have the same set of fixed points, regardless of the vertex permutation $\Pi$ of $\mc{S}'$.

\begin{mybox2}
\begin{observation}\label{obs:samefp}
A \syds{} and an \sds{} with the same underlying graph and the same local functions have the same set of fixed points.
\end{observation}
\end{mybox2}

Since a fixed point of a dynamical system corresponds to an NE of the underlying anti-coordination game, it follows that the complexities of \eqe{}/\eqf{} are
the same for \snsyacg{} and \snsqacg{}.   
The same observation holds for \sesyacg{} and \sesqacg{}.

\subsection{Intractability for the \withmemory{} mode}
\par We establish that \eqe{} (and therefore \eqf{}) is hard for the anti-coordination games under the \se{} mode (i.e., \sesyacg{} and \sesqacg{}). In particular, we present a reduction from \ts{3SAT} to \eqe{} for the \sesyacg{} (modeled as a \syds{}), and by Observation~\ref{obs:samefp}, the hardness is carried over to the \sesqacg{}. Further, the reduction is parsimonious, which further implies that \#\eqe{} is  \textbf{\#P}-hard. {\em We present detailed proofs in the Appendix.}

\begin{mybox2}
\begin{theorem}
For both \sesyacg{} and \sesqacg{}, \eqe{} is \textbf{NP}-complete, and the counting problem \#\eqe{} is \textbf{\#P}-hard. These results hold even when the graph is bipartite.
\end{theorem}
\end{mybox2}

\begin{proofsketch}
Given a \ts{3SAT} formula $f$, we construct a \sesyds{} $\mc{S}$ such that there is a one-to-one correspondence between satisfying assignments of $f$ and fixed points of $\mc{S}$. The construction creates a positive and a negative literal vertex for each variable in $f$, and a clause vertex for each clause. Each clause vertex is adjacent to the corresponding literal vertices. We construct two carefully designed gadgets to ensure that $(i)$ in any fixed point of $\mc{S}$, a positive and a negative literal vertex for the same variable have complementary values; $(ii)$ there is a one-to-one correspondence between satisfying assignments and fixed points.
\end{proofsketch}

\subsection{Finding NE under the \withoutmemory{} mode}
It is known that a \snsqacg{} always has an NE, and an NE can be found in {\em finite} time~\cite{monderer1996potential}. Thus, \eqe{} is trivially true for \snsqacg{}, and by Observation~\ref{obs:samefp}, \eqe{} is also true for \snsyacg{}; however, {\em the complexity of the search problem \eqf{} is \textbf{not} implied.} 

\par In this section, we look beyond the existence problem and show that an NE for an \sn{} game (i.e., \snsyacg{} and \snsqacg{}) can be found in \textit{polynomial} time. 
Specifically, we show that starting an \snsds{} $\mc{S}'$ (modeling a \snsqacg{}) from any configuration $C$, a fixed point of $\mc{S}'$ is always reached in at most $3m$ steps. Since each step of $\mc{S}'$ can be carried out in $O(m)$ time, a fixed point of $\mc{S}'$ can be found in $O(m^2)$ time (Theorem~\ref{thm:sn-conv}). As for a \snsyds{} $\mc{S}$ (modeling a \snsyacg{}), we can transform it into a corresponding \sds{} $\mc{S}'$, and then find a fixed point of $\mc{S}'$ (obtained as described above), which is also a fixed point of $\mc{S}$.  We provide more details below.

\noindent
\textbf{The potentials.} Let $\mc{S}'$ be an \snsds{}. Recall that for each vertex $u \in V(G_{\mc{S}'})$, $\tau_1(u)$ is the minimum number of state-0 neighbors of $u$ such that  $f_u$ evaluates to $1$.
For the \sn{}  mode, $\tau_0(u) = d(u) + 1 - \tau_1(u)$ is the minimum number of state-1 neighbors of $u$ such that  $f_u$ evaluates to $0$. Given a configuration $C$ of $\mc{S}'$, the potentials $\mc{P}$ of vertices, edges, and $C$ are defined as follows.

\smallskip
\noindent
\underline{Vertex potential}: The potential of $u \in V(G_{\mc{S}'})$ under $C$ is $\mc{P}(C, u) = \tau_0(u)$ if $C(u) = 0$; $\mc{P}(C, u) = \tau_1(u)$ otherwise.

\smallskip
\noindent
\underline{Edge potential}: The potential of $e = (u, v) \in E(G_{\mc{S}'})$ under $C$ is $\mc{P}(C, e) = 1$ if $C(u) = C(v)$; $\mc{P}(C, e) = 0$ otherwise.

\smallskip
\noindent
\underline{Configuration potential}: The potential of $C$ is the sum of the vertex potentials and edge potentials over all vertices and edges. That is, $\mc{P}(C, \mc{S}') = \sum_{u \in V(G_{\mc{S}'})} \mc{P}(C, u) + \sum_{e \in E(G_{\mc{S}'})} \mc{P}(C, e)$.

\smallskip
We now establish that under any configuration $C$, the potential is lower and upper bounded by polynomials in $n$ and $m$. A detailed proof is in the Appendix. 

\begin{mybox2}
\begin{lemma}\label{lemma:sds-bound-main}
For any configuration $C$ of $\mc{S}'$, we have 
\begin{equation}
    0 \leq \mc{P}(C, \mc{S}') \leq 3m
\end{equation}
\end{lemma}
\end{mybox2}

\begin{proofsketch}
One can easily verify that the configuration potential over any configuration is at least $0$. As for the upper bound, we argue that
\begin{equation}
    \sum_{u \in V(\sydsG{})} \mc{P}(C, u) \leq \sum_{u \in V(\sydsG{})} \max\{\tau_0(u), \tau_1(u)\} \leq \sum_{u \in V(\sydsG{})} d(u) = 2m
\end{equation}
and that the upper bound for the edge potential is $m$. Thus, the upper bound of $3m$ for the configuration potential follows.
\end{proofsketch}

Lemma~\ref{lemma:sds-bound-main} establishes that the configuration potential gap between any two configurations is at most $3m$ for $\mc{S}'$.

\vspace{-0.5cm}
\paragraph{Decrease of potential.} Next, we argue that whenever a substep of $\mc{S}'$ changes the state of a vertex, the configuration potential decreases by at least~$1$. Consequently, the system reaches a fixed point (i.e., no vertices further update their states) in at most $3m$ total steps. The detailed proof appears in the Appendix.

\begin{mybox2}
\begin{lemma}\label{lemma:decrease-of-potential-main}
Suppose that in a substep of $\mc{S}'$ for a vertex $u$, the evaluation of $f_u$ results in a state change. 
Let $C$ denote the configuration before the substep, and $\hat{C}$ the configuration after the substep.
Then, $\mc{P}(\hat{C}, \mc{S}') - \mc{P}(C, \mc{S}') \leq -1$.
\end{lemma}
\end{mybox2}

\begin{proofsketch}
Since $u$ is the only vertex that undergoes a state change, the overall configuration potential is affected by only the change of $u$'s potential and the potentials of edges incident to $u$. W.l.o.g., suppose $u$'s state changes from $0$ to $1$ in the transition from $C$ to $\hat{C}$. 
We can show that the change in the configuration potential is of the form 
\begin{equation}
    \mc{P}(\hat{C}, \mc{S}') - \mc{P}(C, \mc{S}') = \tau_1(u) + d_1(u) - \tau_0(u) - d_0(u)
\end{equation}
where $d_0(u)$ and $d_1(u)$ are the numbers of $u$'s neighbors in state-$0$ and state-$1$ in $C$, respectively. Since $\hat{C}(u) = 1$, it follows that $d_0(u) \geq \tau_1(u)$ and $d_1(u) \leq \tau_0(u) - 1$. Therefore, $\mc{P}(\hat{C}, \mc{S}') - \mc{P}(C, \mc{S}') \leq -1$.
\end{proofsketch}

\par From the above two Lemmas, starting from an arbitrary configuration of \snsds{} $\mc{S}'$, a fixed point is reached in $3m$ steps. 
Thus, we can find a fixed point by starting with an arbitrary inititial configuration,
and simulating the operation of $\mc{S}'$ for $3m$ steps.
Each step consists of $n$ substeps, each of which evaluates one of the local functions.
Thus, each step can be simulated in $O(m)$ time. Overall, a fixed point of $\mc{S}'$ can be found in $O(m^2)$ time. Similarly, given a \snsyds{} $\cals{}$, we can first convert $\cals{}$ into an \sds{} $\cals{}'$ by assigning a random vertex permutation, and then find a fixed point of $\mc{S}'$, which is also a fixed point $\mc{S}$. Lastly, given that a fixed point of a dynamical system is an NE of the underlying \withoutmemory{} game, we have:
\begin{mybox2}
\begin{theorem}~\label{thm:sn-conv}
For both \snsqacg{} and \snsyacg{}, we can find an NE in $O(m^2)$ time.
\end{theorem} 
\end{mybox2}

\vspace{-0.5cm}
\paragraph{Remark.} A key reason for the drastic contrast in the complexity of \eqe{}/\eqf{} between the \sn{} and \se{} modes is the difference in the behavior of \textit{threshold-1 vertices} (i.e., vertices with  $\tau_1 = 1$). In the \se{} mode, a threshold-1 vertex $u$ \textit{cannot} be in state $0$ under any fixed point because $u$ will change to state 1 in the next step. (Since $u$ counts its own state, there is at least one $0$ input to $f_u$). In particular, the gadgets used in the hardness proof for the \se{} mode critically depend on this fact. In contrast, such a constraint does \textit{not} hold for threshold-1 vertices under the \sn{} mode. Hence, our hardness proof does not carry over to the \sn{} mode.

\subsection{Efficient algorithms for special classes}
Given the hardness of \eqe{}/\eqf{} for \se{} anti-coordination games, we identify several sufficient conditions under which an NE can be obtained efficiently (if one exists), as follows.

\begin{mybox2}
\begin{theorem}
For both \sesyacg{} and \sesqacg{}, there is a $O(m+n)$ time algorithm for \eqe{}/\eqf{} 
for \textbf{any} of the following restricted cases: $(i)$ The underlying graph is a complete graph. $(ii)$ The underlying graph has no even cycles. $(iii)$ The underlying graph is a directed acyclic graph. $(iv)$ The threshold for each $u$ satisfies $\tau_1(u) \in \{1, d_u + 1\}$.
\end{theorem}
\end{mybox2}

\begin{proofsketch}
The detailed proof appears in the appendix. Here, we provide a sketch for $(i)$ (i.e., complete graphs). 
Let $\mc{P} = \{V_1, ..., V_k\}$ be the partition of the vertex set $V$ 
where each block of $\mc{P}$ consists of a maximal set of vertices with the same value of $\tau_1$, 
and the blocks of $\mc{P}$ are indexed so that for each $i$, $1 \leq i < k$,
the members of $V_i$ have a lower value of $\tau_1$ than the members of $V_{i+1}$.
Suppose that configuration $C$ is a fixed point.
Let $q$ be the highest block index such that the number of 0's in $C$ is at least the $\tau_1$ value
of the vertices in $V_q$.
Since the underlying graph is a complete graph, for each vertex $u$,
$C(u) = 1$ iff $u \in V_i$ for some $i$ where $i \leq q$.
Thus there are only $k+1$ candidate configurations that can possibly be a fixed point.
Our algorithm constructs each of these candidates and checks if it is a fixed point.
\end{proofsketch}

\section{Convergence}
We have shown in the previous section that {\em for \snsqacg{}, starting from any action profile, the best-response dynamics converges to an NE in $O(m)$ steps}.
In contrast, it is known that the best-response for \sesqacg{} could have exponentially long limit cycles~\cite{barrett2003some}. 

\vspace{-0.5cm}
\paragraph{Remark.} The dynamics of an \sds{} and its corresponding \syds{} can be drastically different. Therefore, the $O(m)$ convergence time for a \sqacg{} (\sds{}) established in the previous section \textbf{does not} imply an $O(m)$ convergence time for a \syacg{} (\syds{}). In fact, synchronous anti-coordination games are \textbf{not} potential games. Therefore, the approach we used for \sqacg{} does not carry over to the analysis of \syacg{}. Also, both \snsyacg{} and \sesyacg{} can have length-$2$ limit cycles, and there are instances of \sesyacg{} that do not have an NE 
(e.g., the underlying graph is an odd cycle and $\tau_1 = 2$ for all vertices.).






\subsection{Convergence of synchronous games}
\textit{Synchronous} anti-coordination games are \textbf{not} potential games. Therefore, the results on potential games by Monderer and Shapley~(\cite{monderer1996potential}) do not apply. As shown in~\cite{adam2012threshold}, the limit cycles of such a game are of length at most $2$. In this section, we study the convergence time to either an NE or a 2-cycle for 
\snsyacg{} and \sesyacg{}. Using a potential function argument inspired by~\cite{goles2013neural}, in Theorem~\ref{thm:sy-conv} we establish that for both \snsyacg{} and \sesyacg{}, starting from an arbitrary action profile, the best-response dynamics converges in $O(m)$ steps. ({\em A detailed proof of the theorem appears in the Appendix.}) 
Here, we present a proof sketch for the \sn{} mode. 
Let $\cals{} = (\sydsG{}, \mc{F})$ be a \snsyds{} corresponding to a given \snsyacg{}.

\vspace{-0.5cm}
\paragraph{The potentials.} Due to the existence of length-2 limit cycles, our definitions of potentials at each time step account for not only the states of vertices at the current step but also that of the next step. 
For \sn{} mode, let $\tau_0(u) = d(u) + 1 - \tau_1(u)$ be the minimum number of state-1 neighbors of vertex $u$ such that $f_u$ evaluates to $0$.
We henceforth assume that no local function is a constant function (i.e., $1 \leq \tau_1(u) \leq d(u), \; \forall u \in V(\sydsG{})$); a justification is given in the Appendix. 
 Consequently, $1 \leq \tau_0(u) \leq d(u)$.
We define $\tilde{\tau}_0(u) = \tau_0(u) - 1/2$. 
Given an arbitrary configuration $\Ct{}$ of $\mc{S}$, let $\Ct{'}$ be the successor of $\Ct{}$,
and $C''$ the successor of $C'$. 

\smallskip
\noindent
\underline{Vertex potentials.} The potential of a vertex $u$ under $\Ct{}$ is defined by
$$\mc{P}(\Ct{}, u) = [\Ct{}(u) + \Ct{'}(u)] \cdot \tilde{\tau}_0(u)$$

\smallskip
\noindent
\underline{Edge potentials.} The potential of an edge $e = (u, v)$ under $\Ct{}$ is defined by
$$\mc{P}(\Ct{}, e) = \Ct{}(u) \cdot \Ct{'}(v) + \Ct{}(v) \cdot \Ct{'}(u)$$

\smallskip
\noindent
\underline{Configuration potential.} The potential of a configuration $\Ct{}$ is defined by
$$\mc{P}(\Ct{}, \mc{S}) = \sum_{e \in E(\sydsG{})} \mc{P}(\Ct{}, e) - \sum_{u \in V(\sydsG{})} \mc{P}(\Ct{}, u)$$

\noindent
We now establish lower and upper bounds
on the potential under an arbitrary
configuration $C$.

\begin{mybox2}
\begin{lemma}\label{lemma:syds-bound-main}
For any configuration $C$ of a \snsyds{} $\mc{S}$, we have
$$-4m + n ~\leq~ \mc{P}(C, \mc{S}) \leq 0$$
\end{lemma}
\end{mybox2}

\begin{proofsketch}
One can easily verify that the maximum value of the sum of vertex potentials is $4m - n$, which gives the lower bound. 
We now address the upper bound of $0$. 
For each vertex $u$, let $E_u$ be the set of edges incident on $u$, 
and let $\sigma_u = \sum_{u \in V(\sydsG{})} C(v)$.
Note that for an inverted-threshold function $f_u$, 
$C'(u) = 1$ iff $\sigma_u < \tau_0(u)$, i.e., 
iff $\sigma_u \leq \tau_0(u) - 1$.
Let $\beta_u = \sigma_u \cdot C'(u) - \Ct{'}(u) \cdot \Tilde{\tau}_0(u) - \Ct{}(u) \cdot \Tilde{\tau}_0(u)$.

The configuration potential can be restated as:
\noindent
$\mc{P}(\Ct{}, \mc{S}) = \sum_{u \in V(\sydsG{})} \beta_u$.
If $\Ct{'}(u) = 0$, then
$\beta_u = - \Ct{}(u) \cdot \Tilde{\tau}_0(u)$,
which is at most $0$.
If  $\Ct{'}(u) = 1$, then
$\beta_u \leq ({\tau}_0(u) - 1) - \Tilde{\tau}_0(u)$,
which is at most $-1/2$.
This concludes the proof. 
\end{proofsketch}

\vspace{-0.5cm}
\paragraph{Decrease of potential.} We show that from any configuration $\Ct{}$, the potential decreases by at least $1/2$ every step until a fixed point or a 2-cycle is reached. Thus, the dynamics converges in at most $8m-2n$ steps.

\smallskip

\begin{mybox2}
\begin{lemma}\label{lem:syn-1/2}
Let $\Ct{}$ be an arbitrary configuration of $\mc{S}$. 
Let $\Delta(\mc{S}) = \mc{P}(\Ct{'}, \mc{S}) - \mc{P}(\Ct{}, \mc{S})$ denote the change of configuration potential from $\Ct{}$ to $\Ct{'}$.
Then $\Delta(\mc{S}) = 0$ if and only if $\Ct{} = \Ct{''}$, that is $\Ct{}$ is a fixed point or is part of a $2$-cycle $\Ct{} \longleftrightarrow \Ct{'}$. Furthermore, if $\Ct{} \neq \Ct{''}$ (i.e., the dynamics has not converged), then, the configuration potential has decreased by at least $1/2$, i.e., $\Delta(\mc{S}) \leq - 1/2$.
\end{lemma}
\end{mybox2}

\begin{proofsketch}
We define the change of potential for an edge $e = (u, v)$ as $\Delta(e) = \mc{P}(\Ct{'}, e) - \mc{P}(\Ct{}, e)$, and the change of potential for a vertex $u$ as $\Delta(u) = \mc{P}(\Ct{'}, u) - \mc{P}(\Ct{}, u)$. 
Then, 
\begin{align}
    \Delta(\mc{S}) &= \mc{P}(\Ct{'}, \mc{S}) - \mc{P}(\Ct{}, \mc{S}) \nonumber = \sum_{e \in E(\sydsG{})} \Delta(e) - \sum_{u \in V(\sydsG{})} \Delta(u)
\end{align}
Rearranging terms from the definition of potentials, we get

\begin{equation}
    \Delta(e) = \Ct{'}(u) \cdot [\Ct{''}(v) - \Ct{}(v)] + \Ct{'}(v) \cdot [\Ct{''}(u) - \Ct{}(u)]
\end{equation}
and 
\begin{equation}
    \Delta(u) = [\Ct{''}(u) - \Ct{}(u)] \cdot \tilde{\tau}_0(u)
\end{equation}

\smallskip

Now we argue that $\Delta(\mc{S}) = 0$ iff $\Ct{} = \Ct{''}$. First suppose that $\Ct{} = \Ct{''}$, so that for every vertex $u$, $\Ct{}(u) = \Ct{''}(u)$. Consequently, $\Delta(e) = 0$, $\forall e \in E(\sydsG{})$ and $\Delta(u) = 0$, $\forall u \in V(\sydsG{})$.  
Now suppose that $\Ct{} \neq \Ct{''}$. Let $V_{0-1}$ denote the set of vertices $u$ such that
$\Ct{}(u) = 0$ and $\Ct{''}(u) = 1$.
Similarly, let $V_{1-0}$ denote the set of vertices $u$ such that
$\Ct{}(u) = 1$ and $\Ct{''}(u) = 0$.
We establish the following two equalities:

\begin{equation}
    \sum_{u \in V(\sydsG{})} \Delta(u) = \sum_{u \in V_{0-1}} \tilde{\tau}_0(u) - \sum_{u \in V_{1-0}} \tilde{\tau}_0(u)
\end{equation}
and 
\begin{equation}
    \sum_{e \in E(\sydsG{})} = \left(\sum_{u \in V_{0-1}} \sum_{(u,v) \in E_u} \Ct{'}(v)\right) - \left(\sum_{u \in V_{1-0}} \sum_{(u,v) \in E_u} \Ct{'}(v)\right)
\end{equation}

\noindent
\smallskip
where $E_u$ is the set of edges incident on $u$. Recall that $\Delta(\mc{S})$ equals the change in edge potentials minus the change in vertex potentials, so by 
(i) and (ii) above,
\begin{align}
    \small
    \Delta(\mc{S}) &= \sum_{u \in V_{0-1}} \left( \left(\sum_{(u,v) \in E_u} \Ct{'}(v)\right) - \tilde{\tau}_0(u)\right) \nonumber + \sum_{u \in V_{1-0}} \left( \tilde{\tau}_0(u) - \left(\sum_{(u,v) \in E_u} \Ct{'}(v)\right)\right)
\end{align}

\noindent
We argue that if $u \in V_{0-1}$, then:

\begin{equation}
    \sum_{(u,v) \in E_u} \Ct{'}(v) \leq \tau_0(u) - 1 = \tilde{\tau}_0(u) - \frac{1}{2}
\end{equation}

\noindent
and thus
$\left(\sum_{(u,v) \in E_u} \Ct{'}(v)\right) - \tilde{\tau}_0(u) \leq -\frac{1}{2}$. Likewise, if $u \in V_{1-0}$, then:
\begin{equation}
    \sum_{(u,v) \in E_u} \Ct{'}(v) \geq \tau_0(u) = \tilde{\tau}_0(u) + \frac{1}{2}
\end{equation}

\noindent
and thus $\tilde{\tau}_0(u) - \sum_{(u,v) \in E_u} \Ct{'}(v) \leq -\frac{1}{2}$. Since $V_{0-1} \cup V_{1-0} \neq \emptyset{}$, we have
\begin{equation}
\Delta(\mc{S}) \leq - \sum_{u \in V_{0-1}} \frac{1}{2} - \sum_{u \in V_{1-0}} \frac{1}{2} \leq -\frac{1}{2}
\end{equation}
and the lemma follows.
\end{proofsketch}

The above discussion establishes that starting from an arbitrary configuration of $\mc{S}$, the dynamics stabilizes in $O(m)$ steps. Our convergence time results for the \sn{} synchronous anti-coordination games follow immediately. In the Appendix, we also show that the above proof can be easily extended to \sesyacg{}.

\begin{mybox2}
\begin{theorem}\label{thm:sy-conv}
For \snsyacg{} and \sesyacg{}, starting from any initial action profile, the dynamics converges in $O(m)$ steps.
\end{theorem}
\end{mybox2}

\vspace{-0.5cm}
\paragraph{Remark.} The polynomial-time convergence (to either an NE or a 2-cycle) for \sesyacg{} does not contradict the results in the previous section where we showed that determining if a \sesyacg{} has an NE is hard. In particular, despite the fast convergence to a limit cycle, the limit cycle will often be a 2-cycle. 

\section{Experimental Studies}
We conduct experiments to study the contrast between the empirical convergence time of synchronous anti-coordination games under the two modes (\sn{} and \se{}) on networks with varying structures, shown 
in Table~\ref{tab:networks}.
Specifically, \texttt{astroph}, \texttt{google+}
and \texttt{Deezer} are real-world 
social networks ~\cite{rozemberczki2020characteristic,leskovec2012learning,leskovec2007graph}, and synthetic networks from the classes \texttt{Gnp}, scale-free~\cite{barabasi1999emergence}, and small-world~\cite{watts1998collective}. Further, we investigate the contrast in the number of Nash equilibria for small problem instances  under the two modes. 

\par All experiments were conducted on Intel Xeon(R) Linux machines with 64GB of RAM. Source code, documentation, and selected networks appear in the code supplement.

\subsection{Experimental results on convergence}
\par \textbf{Convergence time on networks.} For each network, we randomly generate $200$ copies of threshold assignments,  where $\tau_1(u)$ is chosen uniformly at random in the range $[1, d(u) + 1]$ for the \se{} mode, and in the range $[1, d(u)]$ for the \sn{} mode. This gives us $200$ problem instances for each network. Next, for each instance, we construct random initial configurations using various probabilities $p$ of each vertex having state $0$. In particular, we vary $p = 0.1, 0.3, ..., 0.9$, and for each $p$, we generate $500$ random initial configurations. This gives us $2,500$ initial configurations for each instance and a total of $500,000$ simulations of the dynamics on each network. The average number of steps to converge are shown in Table~\ref{tab:networks}. Note that this average for all the examined networks is less than $20$ for both modes. Further, the maximum number of steps (over all simulations) for the \se{} and \sn{} modes are $53$ and $24$
respectively. 
\begin{center}
\small
 \resizebox{0.7\textwidth}{!}{\begin{tabular}{||l c c c c||} 
 \hline
 \textbf{Network} & $n$ & $m$ & \textbf{Avg. steps (\se{})} & \textbf{Avg. steps (\sn{})}\\ [0.5ex] 
 \hline 
  \texttt{Small-world} & 10,000 & 90,000 &  14.41 & 12.71 \\ \hline
  
 \texttt{Scale-free} & 10,000 & 97,219 & 12.78 & 11.66 \\ \hline
 
 \texttt{Gnp} & 10,000 & 99,562 & 13.09 & 12.71\\ \hline
  
  \texttt{astroph} & 17,903 & 196,972 & 17.31 & 13.72\\ \hline
  
  \texttt{google+} & 23,613 & 39,182 & 7.36 & 7.05\\ \hline
  
  \texttt{Deezer} & 28,281 & 92,752 & 14.41 & 12.29\\ \hline
\end{tabular}}
\captionof{table}{\textbf{Convergence for different networks}. Average number of time steps for the best-response dynamics to converge to a NE or a 2-cycle under the \se{} and \sn{} modes.}
\label{tab:networks}
\end{center}

\begin{figure}[!h]
\small
  \centering
    \includegraphics[width=0.5\textwidth]{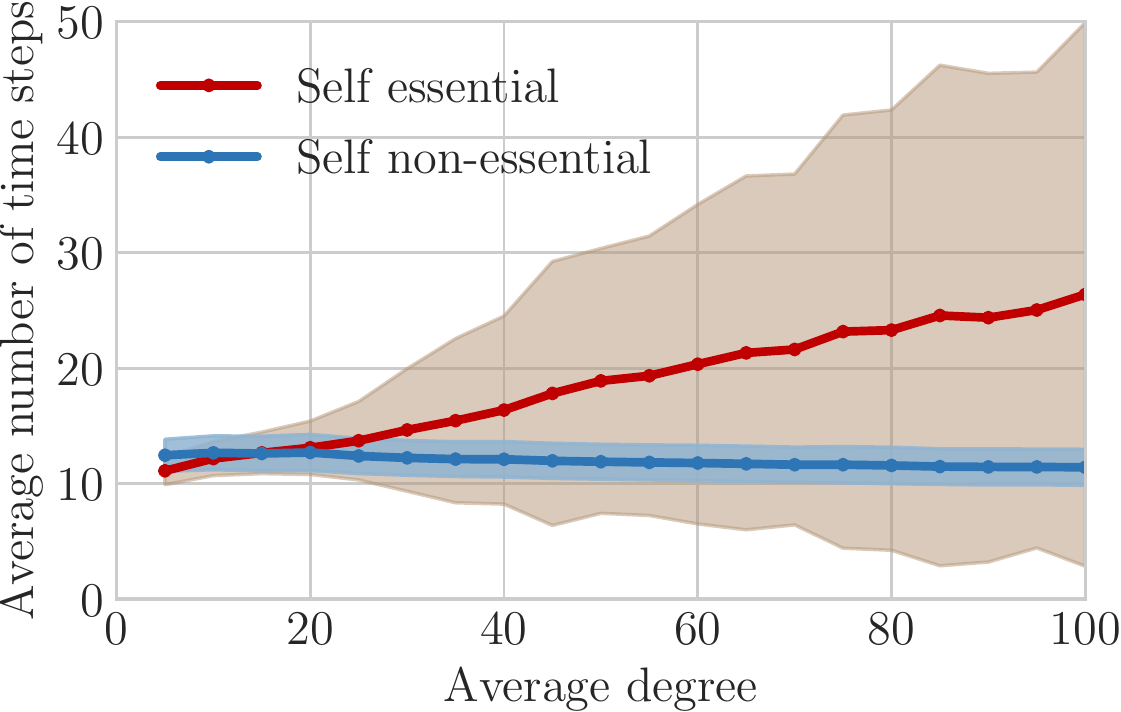}
    \caption{Impact of network density on the average number of steps for the \se{} and \sn{} modes to converge. The underlying \texttt{Gnp} networks have $10,000$ vertices with average degrees varying from $5$ to $100$. 
    The variances for the \se{} and the \sn{} modes are shown in the beige and blue shaded regions, respectively.}
    \label{fig:gnp_avg}
\end{figure}

\noindent
\textbf{\textbf{Impact of network density on convergence time.}} We study the contrast in the average convergence time between the two modes under different network densities. Specifically, we simulate the dynamics on \texttt{Gnp} networks of size $10,000$, with average degrees varying from $5$ to $100$. The results appear in Fig.~\ref{fig:gnp_avg}. The variances of the two modes are shown as shaded regions, with one \textit{stdev} above and below the mean. Overall, as the network density increases, we observe a close to linear increase in the convergence time and the variance for the \se{} mode. In contrast, the convergence time and the variance for the \sn{} mode change marginally. 

\par To gain additional insight, we computed the average (over all pairs of consecutive steps) number $\overline{n}$ of vertices whose states change every $2$ steps until convergence. As suggested by Lemma~\ref{lem:syn-1/2} in section 5, a higher $\overline{n}$ implies a faster decrease in the system potential. Overall, the value of $\overline{n}$ for the \se{} mode and \sn{} mode are $312.24$ and $544.4$, respectively. This provides one reason for the observed difference in the convergence time. Further, the maximum convergence time over all simulations for the \se{} mode is $186$, whereas that for the \sn{} mode is only $25$. 

\subsection{Experimental results on NE existence}
As we have shown, determining whether a game has an NE is intractable for the \se{} mode but easy for the \sn{} mode. Here, we compare the number of NE in small instances between the two modes. In particular, we construct $100$ \texttt{Gnp} networks of size $20$ with average degrees of $4$. For each \texttt{Gnp} network, we construct $200$ different threshold assignments where $\tau_1(u)$ is selected uniformly at random in range $[1, d(u)]$. 
This gives us a total of $20,000$ instances for each mode. Lastly, for each instance, we check all $2^{20}$ possible configurations and count the number of NE among them.

\begin{figure}[!h]
\small
  \centering
    \includegraphics[width=0.5\textwidth]{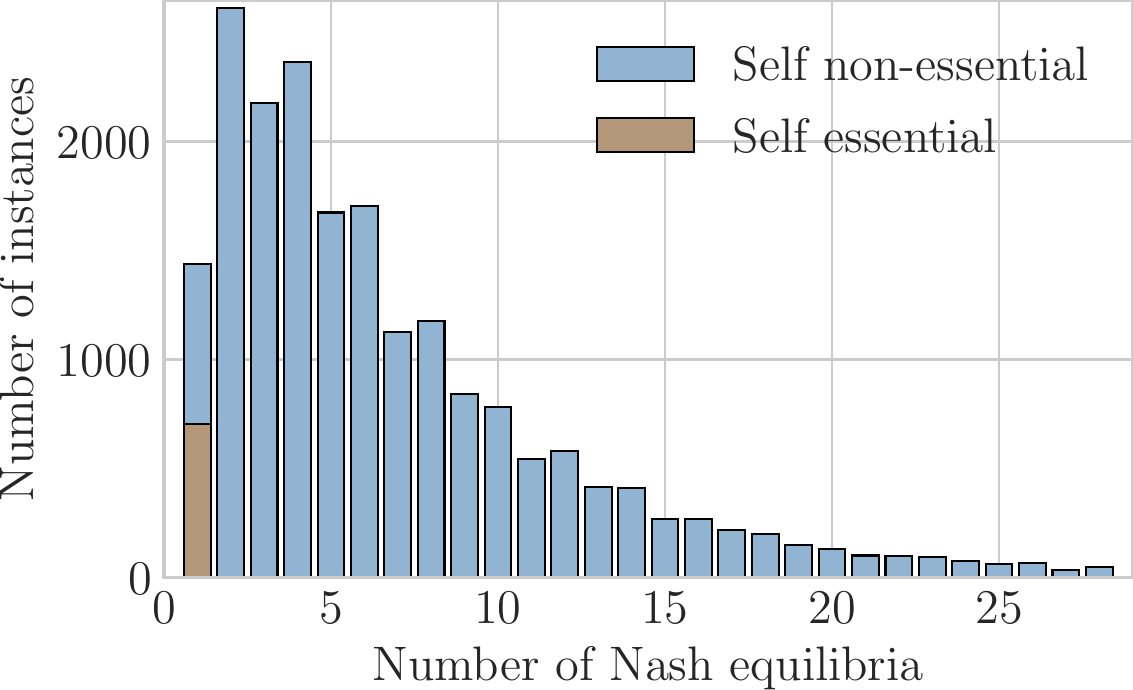}
    \caption{The distribution of the number of instances with at most 28 NE. The underlying \texttt{Gnp} networks are of size $20$ with an average degree of $4$.
    }
    \label{fig:fp_dis}
\end{figure}


\noindent
\textbf{The contrast in the number of NE.} For the \se{} mode, $710$ instances have NE. Among these $710$ instances, $706$ have exactly one NE each, and the remaining $4$ instances have $2$ NE each. In contrast, all the $20,000$ instances for the \sn{} mode have at least one NE, and the average number of NE per instance is $7.26$. Specifically, $75.61\%$ of the \sn{} instances have at most $9$ NE, and $98\%$ have at most $28$ NE. For $1 \leq \eta \leq 28$, the number of instances with $\eta$ NE is shown in Fig.~\ref{fig:fp_dis}. This shows a contrast in the number of NE for the two modes. Further, most configurations that are NE under the \sn{} mode are no longer NE under the \se{} mode.

\section{Conclusions and Future Work}
We studied the problem of finding Nash equilibria in
evolutionary anti-coordination games. We also considered the convergence problem for such games.
Our results present a contrast between the \withmemory{} (\se{}) and \withoutmemory{} (\sn{}) modes w.r.t. the complexity of finding NE. Further, we rigorously established an upper bound on the convergence time for both modes by showing that the best-response dynamics reaches a limit cycle in a polynomial number of steps. One possible future direction is to tighten the bound on convergence time as the empirical convergence time is much smaller than the theoretical bound. Another direction is to study the problem of  finding an {\em approximate} NE~\cite{feder2007approximating} for the \se{} anti-coordination games, since finding an exact NE is hard under the \se{} mode. 
Lastly, it is also of interest to examine the existence of other forms of NE such as mixed-strategy equilibria in anti-coordination games.

\clearpage

\bibliographystyle{plain}
\bibliography{ref}

\newtheorem{innercustomgeneric}{\customgenericname}
\providecommand{\customgenericname}{}
\newcommand{\newcustomtheorem}[2]{%
  \newenvironment{#1}[1]
  {%
   \renewcommand\customgenericname{#2}%
   \renewcommand\theinnercustomgeneric{##1}%
   \innercustomgeneric
  }
  {\endinnercustomgeneric}
}

\newcustomtheorem{customthm}{Theorem}
\newcustomtheorem{customlemma}{Lemma}
\newcustomtheorem{customobs}{Observation}
\newcustomtheorem{customclaim}{Claim}
\newcustomtheorem{customcoro}{Corollary}

\clearpage

\appendix
\begin{center}
\fbox{{\Large\textbf{Appendix}}}
\end{center}

\begin{center}
 \begin{tabular}{||l | l||} 
 \hline
 \textbf{Symbols} & \textbf{Definition}\\ [0.5ex]
 \hline 
 \textsf{SACG} & Sequential anti-coordination games \\ \hline
 \textsf{SyACG} & Synchronous anti-coordination games \\ \hline
 \se{} & Self essential\\ \hline
 \sn{} & Self non-essential\\ \hline
 \sesqacg{} & Sequential anti-coordination games under self essential mode\\ \hline
\snsqacg{} & Sequential anti-coordination games under self non-essential mode\\ \hline
\sesyacg{} & Synchronous anti-coordination games under self essential mode\\ \hline
\snsyacg{} & Synchronous anti-coordination games under self non-essential mode\\ \hline
\sds{} & Sequential dynamical system \\ \hline
SyDS & Synchronous dynamical system \\ \hline
IT & Inverted threshold\\ \hline
\sesds{} & Sequential dynamical system under self essential mode\\ \hline
\snsds{} & Sequential dynamical system under self non-essential mode\\ \hline
\sesyds{} & Synchronous dynamical system under self essential mode\\ \hline
\snsyds{} & Synchronous dynamical system under self non-essential mode\\ \hline
 $\mc{S} = (\sydsG{}, \mc{F})$ & A \syds{} $\mc{S}$ with underlying graph $\sydsG{}$ and set of local functions $\mc{F}$\\ \hline
 $\mc{S'} = (G_{\mc{S}'}, \mc{F}, \Pi)$ & An \sds{} $\mc{S}'$ where $\Pi$ is the vertex update sequence \\ \hline
 $n$ & The number of vertices in $\sydsG{}$ \\ \hline
 $m$ & The number of edges in $\sydsG{}$\\ \hline
 $f_v$ & Local function of $v$\\ \hline
 $\tau_1(v)$ & For vertex $v$, the threshold value on the number of 0's for $f_v$ to equal $1$ \\ \hline
 $\tau_0(v)$ & For vertex $v$, the threshold value on the number of 1's for $f_v$ to equal $0$ \\ \hline
 $N(v)$ & The open neighborhood of $v$, i.e., the set of vertices adjacent to $v$ \\ \hline
 $N^+(v)$ & The closed neighborhood of $v$, i.e., the set of vertices adjacent to $v$ and $v$ itself \\ \hline
 $d(v)$ & The degree of $v$ \\ \hline
 $C$ & A configuration of $\mc{S}$\\ \hline
 $C'$ & The successor of $C$\\ \hline
 $C''$ & The successor of $C'$\\ \hline
 $C(v)$ & The state of $v$ in configuration $C$ of $\mc{S}$\\ \hline
\end{tabular}
\captionof{table}{Symbols and Notation.}
\label{tab:notation}
\end{center}

\paragraph{Remark.} Note that $\tau_1(v)$ uniquely determines $\tau_0(v)$. In particular, $\tau_0(v) + \tau_1(v) = d(v) + 2$ for the \withmemory{} mode, and $\tau_0(v) + \tau_1(v) = d(v) +1$ for the \withoutmemory{} mode. Based on the mapping between anti-coordination games and discrete dynamical systems, we present all proofs in the context of dynamical systems.

\section*{A Detailed Discussion of Related Work}

It is known that the \withoutmemory{} {\em sequential} anti-coordination games are potential games, and by the argument from Monderer and Shapley~\cite{monderer1996potential}, such a game always has an NE.  
In particular, a potential game admits a potential function that increases when an agent chooses an action that yields a strictly higher payoff~\cite{monderer1996potential}. 
Given a potential game starting from an initial action profile, consider the sequence of profiles such that at each step, a player updates its action to obtain a strictly higher payoff (if possible). Such a sequence is called an {\em improvement path}, and for potential games, all the improvement paths are of finite length (known as the {\em finite improvement property}). More importantly, a maximal improvement path (i.e., an improvement path that goes to a maximum of the potential) always ends at an equilibrium point~\cite{monderer1996potential}. Note that this result does not immediately imply that an NE can be reached in polynomial time, as the number of possible action profiles is exponential in the number of agents. Further, we emphasize that anti-coordination games under {\em synchronous} update schemes are \textbf{not} potential games as limit cycles of length $2$ exist. 

The book by Goles and Martinez~\cite{goles2013neural} discussed the phase space properties of dynamical systems, and one can verify that such systems also model {\em coordination} games. In particular, they prove that for synchronous {\em coordination} games, the length of any limit cycle is at most $2$. Their argument uses a mathematical tool called algebraic invariants, and they show that if we consider each limit cycle as a periodical sequence, then the length of such a sequence is either $1$ or $2$. In the same work, Goles and Martinez proposed a Lyapunov function for synchronous {\em coordination games} and show that the best-response dynamics of the game converges to a limit cycle of length at most $2$ in a polynomial number of steps. For anti-coordination games, Barrett et al.~(\cite{barrett2003some}) study phase space properties of sequential dynamical systems (which includes modeling sequential AC games). Their results imply that when local functions are \texttt{nad}'s and \texttt{nor}'s (which are inverted-threshold functions), the length of a limit cycle in a \withmemory{} sequential anti-coordination game can be $2^{O(\sqrt{n})}$ where $n$ is the number of agents. Later, Adam et al.~\cite{adam2012threshold} use a combinatorial approach and argue that the length of a limit cycle in a synchronous anti-coordination game is at most $2$. However, they did not bound the convergence time to reach a limit cycle. 

A more recent work by Ramazi, Riehl, and Cao~\cite{ramazi2016networks} study the convergence time for asynchronous \withoutmemory{} anti-coordination games. In their asynchronous dynamics, agents are updated at each step in a random order, and for each agent, the number of steps between any two consecutive updates is guaranteed to be finite. Based on this scheme, they show that under the best-response dynamics, an equilibrium is always reached in finite time.

  
 

\section*{Additional Material for Section 4}
In this section, we present the detailed proofs of the results given in section 4 of the main manuscript. We start with a key observation: 

\begin{mybox2}
\begin{customobs}{4.1}\label{obs:eqesameclass}
A \syds{} and an \sds{} with the same underlying graph and the same set of local functions have the same set of fixed points.
\end{customobs} 
\end{mybox2}

\noindent
Consequently, \snsyacg{} (\sesyacg{}) and \snsqacg{} (\sesqacg{}) have the same complexity for \eqe{} / \eqf{}.

\section*{4.1 \;\; Intractability for the \withmemory{} mode}
We establish that \ts{Equilibrium existence} (\eqe{}) is \textbf{NP}-hard for \textit{self essential} \textit{synchronous} anti-coordination games (\sesyacg{}), and the problem remains hard on bipartite graph. Further, the corresponding counting problem is \textbf{\#P}-complete. This immediately implies that \ts{Equilibrium finding} (\eqf{}) is also hard for \sesyacg{}. In particular, Let $\cals$ be a \sesyds{} that models a \sesyacg{}. We show that determining the existence of a fixed point of $\cals$ is intractable. By Observation~\ref{obs:eqesameclass}, it follows that \eqe{} / \eqf{} is also hard  for \sesqacg{}. We now proceed with the proof.

\begin{mybox2}
\begin{customobs}{4.2}\label{obs:threshold_1}
Suppose that a vertex $v$ of a \sesyds{} $\mc{S}$ has threshold $\tau_1(v) = 1$. Then in every fixed point (if any exists) $C$ of $\mc{S}$, $C(v) = 1$ and at least one neighbor of $v$ has state $0$ in $C$.
\end{customobs}
\end{mybox2}

\begin{mybox2}
\begin{customobs}{4.3}\label{obs:threshold_degree}
Suppose that a vertex $v$ of a \sesyds{} $\mc{S}$ has threshold $\tau_1(v) = d(v)$. Then in every fixed point (if any exists) $C$ of $\mc{S}$, if $C(v) = 1$, then then all neighbors of $v$ have state 0 in $C$, and if $C(v) = 0$, then at least two neighbors of $v$ have state $1$ in $C$.
\end{customobs}
\end{mybox2}

\begin{mybox2}
\begin{customlemma}{4.4}\label{lemma:gadget}
Let \cals{} be a \sesyds{} whose underlying graph $\sydsG{}$ is a complete bipartite graph of the form:
\begin{enumerate}
    \item[a.] The bipartitions $V(\sydsG{}) = \{A, B\}$.
    \item[b.] $|A| = |B| = 3$.
    \item[c.] $\tau_1(v) = 3, \; \forall v \in V(\sydsG{})$.
\end{enumerate}
The phase space of $\mc{S}$ has only $2$ distinct fixed points for which either $(i)$ vertices in $A$ are in state $1$, and vertices in $B$ are in state $0$, or $(ii)$ vice versa.
\end{customlemma} 
\end{mybox2}

\begin{figure}[!h]
  \centering
    \includegraphics[width=0.5\textwidth]{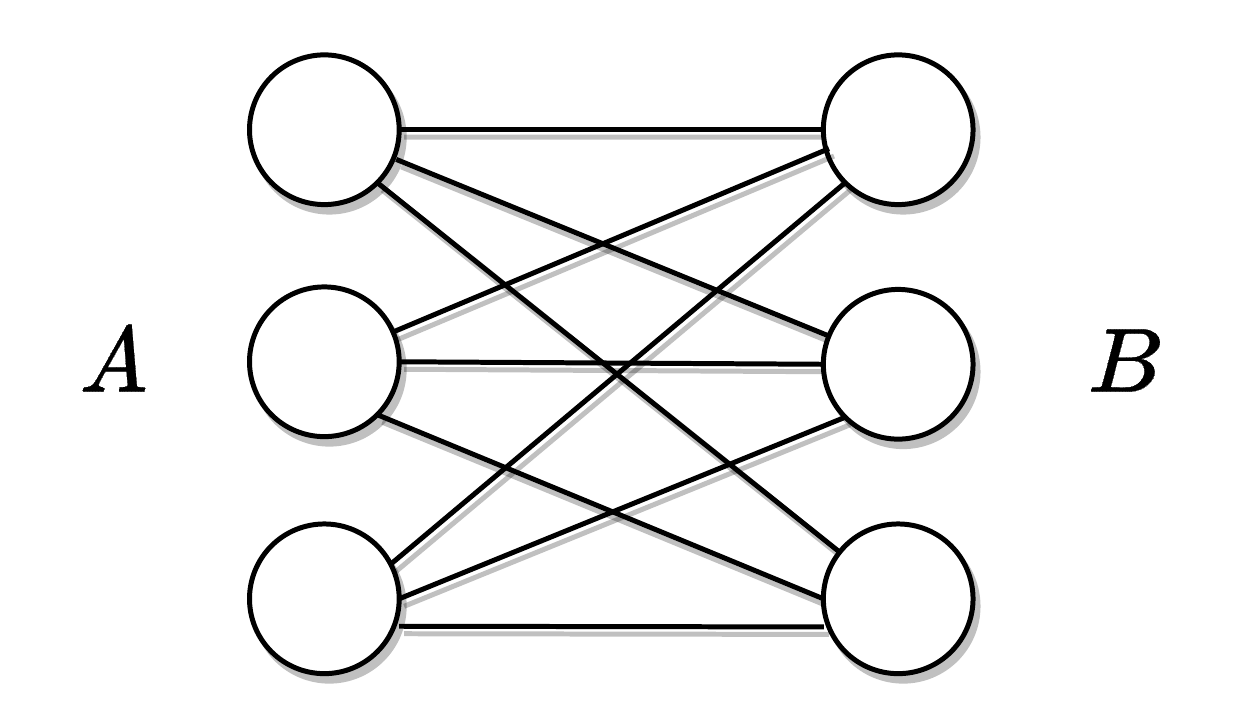}
    \caption{An example \sesyds{} \cals{} for Lemma~\ref{lemma:gadget}, where $\sydsG{}$ is a complete bipartite graph with bipartitions $\{A, B\}$. The thresholds $\tau_1$ of all vertices are $3$.}  
    \label{fig:gadget}
\end{figure}

\begin{proof}
By the threshold values of vertices in $A$ and in $B$, it is easy to see that the two configurations given in the Lemma are fixed points. We now argue that the system has no other fixed points. Let $C$ be a configuration of $\mc{S}$ that is different from the two fixed points above. Let $C'$ be the successor of $C$. We first consider the case where both $A$ and $B$ have at least one state-$1$ vertex under $C$. Let $v \in A$ be such a vertex in state $1$ under $C$. Note that since $B$ has at least one state-$1$ vertex, the number of state-$0$ vertices in $v$'s closed neighborhood is at most $|B| - 1 < \tau_1(v)$. Thus, we have $C'(v) = 0$ and $C$ is not a fixed point. Next, we consider the case where both $A$ and $B$ have at least one state-$0$ vertex under $C$. Let $v$ and $w$ be two state-$0$ vertices in $A$ and $B$, respectively. Since $\tau_1(u) = 3$, $\forall u \in V(\sydsG{})$, $C(v) = 0$ implies that there exists at least two state-$1$ vertex in $B$. Similarly, $C(w) = 0$ implies that there exists at least two state-$1$ vertex in $A$. By our previous argument, it follows that $C$ is not a fixed point. This concludes the proof.
\end{proof}

\noindent
We now present the Theorem on the intractability of \eqe{} for \sesyacg{}.

\begin{mybox2}
\begin{customthm}{4.2}
For \sesyacg{}, the \ts{Equilibrium existence}(\eqe{}) is \textbf{NP}-complete and the counting problem \#\eqe{} is \textbf{\#P}-complete. Further, the problem remains hard on bipartite graphs.
\end{customthm}
\end{mybox2}

\begin{proof}
One can easily verify that the problem is in \textbf{NP}. We now establish the \textbf{NP}-hardness of \eqe{} via a reduction from \ts{3SAT}.
Let $f$ be the formula for an arbitrary \ts{3SAT} instance. We construct a \syds{} $\mc{S}$ such that $\mc{S}$ has a vertex for each literal, a vertex for each clause, and a collection of gadgets. Further, $\mc{S}$ is constructed so that there is a one-to-one correspondence between fixed points of $\mc{S}$ and satisfying assignments to the given \ts{3SAT} formula $f$.

\par \textit{Literal vertices.} For each variable $x_i$ in $f$, there is a positive literal vertex $y_i$ and a negative literal vertex $z_i$ in $\mc{S}$. Both these vertices have threshold $\tau_1(y_i) = \tau_1(z_i) = 3$. Under any configuration, we interpret positive literal vertex $y_i$ having state 0 
as corresponding to variable $x_i$ being true, and negative literal vertex $z_i$ having value 0 as corresponding to variable $x_i$ being false. Our Gadgets (shown later) ensure that in any fixed point of $\mc{S}$, these two vertices have complementary values.

\textit{Clause vertices.} For each clause $c_j$ in $f$, there is a clause vertex $w_j$, $\tau_1(w_j) = 1$. Further, $w_j$ is adjacent to each of the literal vertices that correspond to literals occurring in clause $c_j$. There is no other edges incident on $w_j$. From Observation \ref{obs:threshold_1}, in every fixed point of $\mc{S}$, $w_j$ has state 1, and at least one of the adjacent literal vertices has state 0.

\textit{Gadgets.} We now construct the gadgets, which involve auxiliary vertices for the literal vertices, as follows. For each variable $x_i$ in $f$, there are three auxiliary vertices, namely $a_i$, $d_i$ and $e_i$. In particular, 

\begin{enumerate}
    \item[$(i)$] The vertex $a_i$ is adjacent to both $x_i$'s positive literal vertex $y_i$ and negative literal vertex $z_i$. Further, $\tau_1(a_i) = 1$. From Observation \ref{obs:threshold_1}, in every fixed point of $\mc{S}$, $a_i$ has state 1, and at least one of the two literal vertices $y_i$ and $z_i$ has state 0.
    \item[$(ii)$] The vertex $d_i$ and $e_i$ has threshold $\tau_1(d_i) = 1$ and $\tau_1(e_i) = 3$, respectively. Further, we introduce three auxiliary edges $(e_i, d_i)$, $(e_i, y_i)$ and $(e_i, z_i)$. From Observation \ref{obs:threshold_1}, in every fixed point, $d_i$ has state 1, thus, its neighbor $e_i$ has state 0. Then, from an application of Observation \ref{obs:threshold_degree} to vertex $e_i$, at least one of the two literal vertices $y_i$ and $z_i$ has state 1.
\end{enumerate}
\par The combined effect of the auxiliary vertices and edges described thus far is that under every fixed point of $\mc{S}$, exactly one of $y_i$ and $z_i$ has state 0, and exactly one has state 1.

Finally, for every literal vertex (positive or negative), denoted by $v$, we add an auxiliary structure based on
the bipartite graph in Lemma~\ref{lemma:gadget}.
The structure involves five auxiliary vertices:
$g_v^1$, $g_v^2$, $h_v^1$, $h_v^2$, and $h_v^3$, each with threshold $\tau_1 = 3$
(which is the degree of each of these vertices).
Consider the subsets $A_v = \{v, h_v^1, h_v^2\}$ and $B_v = \{g_v^1, g_v^2, g_v^3\}$. The structure has the following nine auxiliary edges: $(v, h_v^1)$, $(v, h_v^2)$, $(v, h_v^3)$, 
$(g_v^1, h_v^1)$, $(g_v^1, h_v^2)$, $g_v^1, h_v^3)$, 
$(g_v^2, h_v^1)$, $(g_v^2, h_v^2)$, $(g_v^2, h_v^3)$. That is, the subgraph of $\mc{S}$ induced on $A_v \cup B_v$ is a complete bipartite graph with bipartitions $A_v$ and $B_v$.

By Lemma~\ref{lemma:gadget},
that in any fixed point of $\mc{S}$,
either all the vertices in $A_v$ have state $1$ and all the vertices in $B_v$ have state 0, or vice versa. To see this, first suppose that in a given fixed point $C$, at least one of the vertices in $B_v$ has state 1. Then, from Observation \ref{obs:threshold_degree},
the three neighbors of this vertex, namely the three vertices in $A_v$, all have state 0. Consequently, since the threshold of each vertex in $B_v$ is 3, all three vertices in $B_v$ have value~1.
Now suppose that none of the vertices in $B_v$ has value~1, i.e., they all have value 0. Since the threshold of each vertex in $A_v$ is 3, all three vertices in $A_v$ have value~1.

This completes the construction of $\mc{S}$ which clearly takes polynomial time. Further, the resulting graph is bipartite. We now claim that there is a one-to-one correspondence between fixed points of $\mc{S}$ and satisfying assignments of $f$.

\begin{itemize}
    \item[$(\Rightarrow)$] Let $\alpha$ be a satisfying assignment of $f$. We construct a configuration $C_{\alpha}$ of $\mc{S}$ as follows. For each variable under $\alpha$, if $\alpha[x_i] = \ts{true}$, then the positive literal vertex $y_i$ has the state $C_{\alpha}(y_i) = 0$, and the negative literal vertex $z_i$ has the state  $C_{\alpha}(z_i) = 1$. On the other hand, if If $\alpha[x_i] = \ts{false}$, then $C_{\alpha}(y_i) = 1$ and $C_{\alpha}(z_i) = 0$. Further, every clause vertex $w_j$ has state 1. As for the states of auxiliary vertices for each $x_i$, we set $a_i$ and $d_i$ to state 1, and $e_i$ to state $0$. Lastly, for each literal vertex $v$, the other two members of $A_v$ have the same value as $v$, and the three members of $B_v$ have the complement of the value of $v$ under $C$. This completes the specification of $C_{\alpha}$. By checking the number of state-$0$ vertices in the closed neighborhood of each vertex, it can be verified that $C_{\alpha}$ is a fixed point of $\mc{S}$.
    
    \item[$(\Leftarrow)$] Let let $C$ be a fixed point of $\mc{S}$. Let $\alpha_C$ be the assignment to $f$ where $\alpha_C(x_i) = 1$ iff $C_{\alpha}[y_i] = 0$. Since $C$ is a fixed point, every clause vertex is adjacent to at least one literal vertex with the value 0. Thus, $\alpha_C$ is a satisfying assignment of $f$.
\end{itemize}

We now have established that determining if $\mc{S}$ has a fixed point is \textbf{NP}-complete, even when the underlying graph is bipartite. Further, it can be verified that $C = C_{\alpha_C}$, so the reduction is parsimonious. The \textbf{NP}-hardness of \eqe{} and the \textbf{\#P}-hardness of the counting version \#\eqe{} for \sesyacg{} immediately follows. This concludes the proof.
\end{proof}

\begin{mybox2}
\begin{customcoro}{{4.5}}
For \sesqacg{}, the \ts{Equilibrium existence}(\eqe{}) is \textbf{NP}-complete and the counting problem \#\eqe{} is \textbf{\#P}-hard. Further, the problem remains hard on bipartite graphs.
\end{customcoro}
\end{mybox2}

\vspace{-0.5cm}
\paragraph{Remark.} A fixed point under the \se{} mode is generally \textbf{not} a fixed point under the \sn{} mode. Therefore, the above hardness result does not imply a hardness of \eqe{} for the \sn{} anti-coordination games.

\section*{4.2 \;\; Finding NE under the \withoutmemory{} mode}
As pointed out in~\cite{vanelli2020games}, a \withoutmemory{} \textit{sequential} anti-coordination game (\snsqacg{}) always has a pure Nash equilibrium (NE). By Observation~\ref{obs:eqesameclass}, a \withoutmemory{} \textit{synchronous} anti-coordination game (\snsyacg{}) also always has an NE. In this section, we explore beyond the existence problem and tackle the problem of finding an NE in a for \sn{} mode under arbitrary network topology (i.e., \eqf{}). In particular, inspired by the potential function approach developed in~\cite{barrett2006complexity}, given a \snsds{} $\mc{S}'$ (modeling a \snsqacg{}), we show that starting from an arbitrary configuration, a fixed point of $\mc{S}'$ is always reached in at most $3m - n$ steps. Since each step of $\mc{S}'$ can be carried out in $O(m)$ time, a fixed point of $\mc{S}'$ is then found in $O(m^2)$ time. Consequently, a fixed point of a \snsyds{} $\mc{S}$ (modeling a \snsyacg{}) can also be founded in $O(m^2)$ time.

\par We note that our results does not follow from~\cite{barrett2006complexity} since $(i)$ we study anti-coordination games, and the results in~\cite{barrett2006complexity} are for \textit{coordination} games; $(ii)$ in the \withoutmemory{} mode, each vertex does not consider its own state while playing the game, whereas~\cite{barrett2006complexity} focuses on the case where its own state is considered by each vertex. If $\tau_1(v) = 0$ or $\tau_1(v) = d_v + 1$ for some vertex $v$, then $v$ is a vertex whose state is constant after at most one transition. 
Thus, we can remove $v$ from the graph and update the threshold of neighbors correspondingly without affecting system dynamics. Without loss of generality, we assume that there are no constant vertices, that is $1 \leq \tau_1(v) \leq d(v)$, $\forall v \in V(G_{\mc{S}'})$. 

\subsection*{The potential functions and bounds}
\par Let $\cals' = (\sdsG{}, \mc{F}', \Pi)$ be a \snsds{} that models a \snsqacg{}. Given a configuration $C$, We now define the potentials of vertices, edges, and the system under $C$. 

\noindent
\underline{The vertex potential}. Given a vertex $u \in V(\sdsG{})$, the potential of $u$ under $C$ is defined as follows
\begin{gather*}
\mc{P}(C, u) =
\begin{cases}
   \tau_0(u) & \text{if }C(u) = 0\\    
    \tau_1(u) & \text{if }C(u) = 1\\ 
\end{cases}
\end{gather*}

\noindent
\underline{The edge potential}. Given an edge $e = (u, v) \in E(\sdsG{})$, the potential of $e$ under $C$ is defined as follows 
\begin{gather*}
\mc{P}(C, e) =
\begin{cases}
   1 & \text{if } C(u) = C(v)\\   
    0 & \text{if }C(u) \neq C(v)\\
\end{cases}
\end{gather*}

\noindent
\underline{The configuration potential}. The potential of the system $\mc{S}'$ under $C$ is defined as the sum of the vertex potentials and edge potentials over all vertices and edges.
$$
\mc{P}(C, \mc{S}') = \sum_{u \in V(\sdsG{})} \mc{P}(C, u) + \sum_{e \in E(\sdsG{})} \mc{P}(C, e)
$$

\vspace{-0.5cm}
\paragraph{A lower bound on the configuration potential.} We first establish a lower bound of the $\mc{P}(C, \mc{S}')$ for any configuration $C$. 

\begin{mybox2}
\begin{customlemma}{{4.3}}\label{lemma:lower-bound}
For any configuration $C$ of $\mc{S}'$, we have 

$$\mc{P}(C, \mc{S}') \geq \left(\sum_{u \in V(\sdsG{})} \min\{\tau_0(u), \tau_1(u)\}\right) + \gamma$$
where $\gamma$ is the minimum number of edges whose endpoints have the same color in $\sdsG{}$, over all 2-coloring of $V(\sdsG{})$.
\end{customlemma}
\end{mybox2}

\begin{proof}
Given any configuration $C$, since the potential of each vertex is either $\tau_1(u)$ or $\tau_0(u)$, it immediately follows that $\sum_{u \in V(\sydsG{})} \min\{\tau_0(u), \tau_1(u)\}$ is the lower bound on the sum of potentials from all vertices. 

\par As for the bound on the sum of edge potentials, remark that $C$ corresponds to a 2-coloring of $V(\sdsG{})$ (i.e., each state represents a color). Thus, the minimum number of edges whose endpoints have the same state under $C$ is at most $\gamma$. Since each edge with the same-state endpoints has potential value $1$, it follows that the sum of edge potential under any configuration $C$ is at least $\gamma$. This concludes the proof.
\end{proof}

\vspace{-0.5cm}
\paragraph{An upper bound on the configuration potential.} Next, we present the upper bound of the configuration potential. Given an arbitrary configuration $C$

\begin{mybox2}
\begin{customlemma}{{4.4}}\label{lemma:upper-bound}
The configuration potential satisfies 
$$\mc{P}(C, \mc{S}') \leq 3m$$
\end{customlemma}
\end{mybox2}

\begin{proof}
Observe that under an arbitrary configuration of $\mc{S}'$, the sum of vertex potential satisfies
$$\sum_{u \in V(\sdsG{})} \mc{P}(C, u) \leq \sum_{u \in V(\sdsG{})} \max\{\tau_0(u), \tau_1(u)\} \leq \sum_{u \in V(\sdsG{})} d(u) = 2m$$

As for the sum of edge potential $\sum_{e \in E(\sdsG{})} \mc{P}(C, e)$, it is easy to see that the upper bound is $m$. The upper bound of the overall configuration potential follows immediately.
\end{proof}

Overall, We have shown that the gap in the configuration potential between an arbitrary initial configuration and a system's converged configuration is at most
\begin{equation}
    3m - \sum_{v \in V(\sydsG{})} \left(\min\{\tau_0(v), \tau_1(v)\}\right) - \gamma \leq 3m - n
\end{equation}

\subsection*{Decrease of the potential after each update}
We establish that starting from an arbitrary initial configuration that is not a fixed point, the configuration potential of $\mc{S}'$ decreases by at least $1$ after each vertex switches its state. Given that the potential gap is at most $3m - n/2$, it follows that the system reaches a fixed point in at most $3m - n/2$ time steps.

\begin{mybox2}
\begin{customlemma}{{4.5}}\label{lemma:decrease-of-potential}
Given a configuration $C$ of $\mc{S}'$ that is not a fixed point. Let $\Tilde{C}$ be a configuration that results from the state change of a single vertex $u$ due to the dynamics of $\mc{S}'$. We then have $\mc{P}(C, \mc{S}') - \mc{P}(\Tilde{C}, \mc{S}') \geq 1$.
\end{customlemma}
\end{mybox2}

\begin{proof}
Since $u$ is the only vertex that undergoes the state change, the overall configuration potential is affected by only the change of $u$'s potential and the potentials of edges incident to $u$. Let $d_0(u)$ and $d_1(u)$ denote the number of $u$'s neighbors in state-$0$ and the number of $u$'s neighbors in state-$1$ under $C$, respectively. Without loss of generality, suppose $u$ changes its state from $C(u) = 0$ to $\Tilde{C}(u) = 1$. Subsequently, the sums of potentials of $u$ and edges incident to $u$ under $C$ and $\Tilde{C}$ are
$$
\mc{P}(C, u) + \sum_{e = (u, v), v \in N(u)} \mc{P}(C, e) = \tau_0(u) + d_0(u)
$$
and
$$
\mc{P}(\Tilde{C}, u) + \sum_{e = (u, v), v \in N(u)} \mc{P}(\Tilde{C}, e) = \tau_1(u) + d_1(u)
$$
respectively. Since $\Tilde{C}(u) = 1$, it follows that $d_0(u) \geq \tau_1(u)$ and $d_1(u) \leq \tau_0(u) - 1$. Therefore, we have 
\begin{align*}
    \mc{P}(C, \mc{S}') - \mc{P}(\Tilde{C}, \mc{S}') &= \mc{P}(C, u) + \sum_{e = (u, v), v \in N(u)} \mc{P}(C, e) -  \left(\mc{P}(\Tilde{C}, u) + \sum_{e = (u, v), v \in N(u)} \mc{P}(\Tilde{C}, e)\right)\\
    &= \tau_0(u) + d_0(u) - \tau_1(u) - d_1(u)\\
    &\geq 1
\end{align*}
This concludes the proof.
\end{proof}

In summary, we have shown that the potential gap between any two configurations is at most $3m - n$. Further, each change of vertex state due to the system dynamic decreases the overall potential by at least one. Since in each time step (before reaching a fixed point), at least one vertex updates its state,  Lemma~\ref{lemma:sequential-reach} immediately follows.

\begin{mybox2}
\begin{customlemma}{{4.6}}\label{lemma:sequential-reach}
For \snsds{}, starting from an arbitrary initial configuration, the system dynamics reaches a fixed point in at most $3m-n$ time steps.
\end{customlemma}
\end{mybox2}

In each time step, we compute the local functions of all vertices to determine the successor configuration, which takes $O(m)$ time. Therefore, a fixed point of  $\cals{}'$ can be obtained in $O(m^2)$ time. Given that a fixed point of a \snsds{} $\cals{}'$ is also a fixed point of its twin \snsyds{} $\cals{}$, we also establish the tractability of finding an NE for \snsyds{}. Our results for \sn{} anti-coordination games follow.

\begin{mybox2}
\begin{customthm}{{4.7}}
For both \snsyacg{} and \snsqacg{}, we can find a pure Nash equilibrium of the game in $O(m^2)$ time.
\end{customthm}
\end{mybox2}

\section*{4.3 \;\; Finding Nash equilibria under special cases}
We have established the intractability of \eqe{} / \eqf{} for \withmemory{} (\se{}) anti-coordination games, whereas a \withoutmemory{} (\sn{}) game admits a polynomial time algorithm for finding an NE. In this section, we identify several special classes of the problem instances such that an NE (if any) can be found in polynomial time for \se{} anti-coordination games. Further, We extend the results of some special classes to \sn{} anti-coordination games such that an NE can be found in linear time. Based on the connection between anti-coordination games and dynamical systems, we present all proofs in the context of synchronous dynamical systems (and the results for sequential systems follow).

\subsection*{Inclination for one action over another}
We consider the special case where agents are inclined to choose one action over the other. Specifically, during the game evolution, each agent $u$ either $(i)$ chooses the action $1$ in the next time step if at least one agent in $u$'s open/closed neighborhood chose action $0$ in the previous time step; $(ii)$ chooses the action $0$ in the next time step if at least one agent in $u$'s open/closed neighborhood chose action $1$ in the previous time step. 

\par Observe that the case $(i)$ above corresponds to $\tau_1(u) = 1$. Further, the case $(ii)$ implies $\tau_1(u) = d(v)$ for the \sn{} game, and $\tau_1(u) = d(v) + 1$ for the \se{} game. We call the local function of a vertex $u$ with $\tau_1(u) = 1$ a \ts{nand} function, and of $u$ with $\tau_1(u) = d(v)$ / $\tau_1(u) = d(v) + 1$ a \ts{nor} vertex function. 

\begin{mybox2}
\begin{customthm}{{4.8}}\label{thm:spacial-nand-nor-withmemory}
For \se{} anti-coordination games and \sn{} anti-coordination games, an NE can be found in $O(m+n)$ time if the corresponding local functions of vertices are \ts{nand}'s and \ts{nor}'s.
\end{customthm}
\end{mybox2}

\begin{proof}
We first present the result for \sesyacg{}, modeled by a \sesyds{} $\cals{} = (\sydsG{}, \mc{F})$. Observe that under any fixed point of $\mathcal{S}$, a \ts{nand} vertex must be in state 1, and a \ts{nor} vertex must be in state 0. This uniquely determines a configuration $C$. We can then compute $C$'s successor $C'$, and examine if $C$ is a fixed point in $O(m + n)$ time. We further establish the claim below

\begin{customclaim}{{4.8.1}}\label{claim:nand-nor-memoryful}
Such a configuration $C$ is a fixed point \textit{if and only if} each \ts{nor} vertex is adjacent to at least one \ts{nand} vertex and vise versa.
\end{customclaim}

For sufficiency, suppose $C$ is a fixed point. If there exists a \ts{nor} vertex $v$ whose neighbors are all \ts{nor} vertices. Then the number of state-0 vertices in $v$' closed neighborhood is $d(v) + 1$ (since $C$ is a fixed point, all \ts{nor} vertices are in state $0$) which equals to the threshold of $v$. Thus, $C(v) = 0 \neq C'(v) = 1$ and $C$ is not a fixed point. An analogous argument applies to the case where $v$ is a \ts{nand} vertex whose neighbors are all \ts{nand} vertices. Specifically, the number of state-$0$ vertices in $v$'s closed neighborhood is $0$. Thus, $C(v) = 1 \neq C'(v) = 0$. As for the necessity of the condition, observe that if a \ts{nor} vertex $v$ is adjacent to at least one \ts{nand} vertex, the number of state-$0$ vertices in $v$'s closed neighborhood is at most $d(v) < \tau_1(v)$, thus, $C(v) = C'(v) = 0$. Similarly, if a \ts{nand} vertex $v$ is adjacent to at least one \ts{nor} vertex, then the number of state-$0$ vertices in $v$'s closed neighborhood is at least $\tau_1(v) = 1$, thus, $C(v) = C'(v) = 1$. This concludes the proof for \sesyacg{} (and thus also \sesqacg{}).

\par We now establish the result for a \snsyacg{}, modeled by a \snsyds{}, $\Bar{\mc{S}} = (G_{\Bar{\mc{S}}}, \mc{F})$. Overall, we first construct a candidate configuration $C$ and then modify $C$ to make it a fixed point of $\mc{S}$. In particular, for each vertex $v \in V(G_{\Bar{\mc{S}}})$, if $v$ is a \ts{nand} vertex (i.e., $\tau_1(v) = 1$), we set $C(v) = 1$. On the other hand, if $v$ is a \ts{nor} vertex (i.e., $\tau_1(v) = d(v)$), set $C(v) = 0$. Based on the same argument in Claim~\ref{claim:nand-nor-memoryful}, it follows that $C$ is a fixed point of $\mathcal{S}'$ if and only if each \ts{nand} vertex is adjacent to at least one \ts{nor} vertex and visa versa. If this condition does not hold, however, there must exist at least one \ts{nand} (\ts{nor}) vertex whose neighbors are all \ts{nand} (\ts{nor}) vertices. Subsequently, We further modify $C$ as follows. First, for each \ts{nand} vertex $v$ whose neighbors are all in state $1$ under $C$, we set $C(v) = 0$. Let $V_{nand}$ denote such a set of vertices. Further, for each \ts{nor} vertex $v$ whose neighbors are all in state $0$ under $C$, we set $C(v) = 1$. Let $V_{nor}$ denote this set of vertices. The pseudocode is shown in Algorithm~\ref{alg:eqe-nand-nor-memoryless}.

\par We now argue that the resulting configuration $C$ is a fixed point. Let $C'$ be the successor of $C$. First, observe that for each \ts{nand} vertex $u \in V(\sydsG{}) \setminus (V_{nand} \cup V_{nor})$, we have $C(u) = 1$ (since we never change the state of vertices that are not in $V_{nand}$ or $V_{nor}$). Furthermore, $u$ must be adjacent to at least one state-$0$ vertex, or else $u$ will be in $V_{nand}$. Similarly, for each \ts{nor} vertex $u \in V(\sydsG{}) \setminus (V_{nand} \cup V_{nor})$, we have that $C(u) = 0$ and $u$ is adjacent to at least one state-$1$ vertex. It immediately follows that $C(u) = C'(u), \; \forall u \in V(\sydsG{}) \setminus (V_{nand} \cup V_{nor})$. We now consider the states of vertices in $V_{nand}$ or $V_{nor}$. For each vertex $v \in V_{nand}$, observe that neighbors of $v$ must all be \ts{nand} vertices, or else, $v$ is adjacent to at least one state-0 vertex (i.e., a $\ts{nor}$ vertex) which contradicts the fact that $v \in V_{nand}$. Furthermore, we claim that 

\begin{customclaim}{{4.8.2}}\label{claim:nand-nor}
Each neighbor of $v$ is not in $V_{nand}$, $\forall v \in V_{nand}$.
\end{customclaim}

For contradiction, suppose $v'$ is a neighbor of $v$ who is also in $V_{nand}$. If $v'$ is added to $V_{nand}$ before $v$, then $v$ cannot also be in $V_{nand}$ since $v$ has $v'$ as a state-$0$ neighbor under $C$. Similarly, if $v$ is added to $V_{nand}$ before $v'$, then $v'$ cannot also be in $V_{nand}$. This concludes the claim. Claim~\ref{claim:nand-nor} implies that $v$ has no state-$0$ neighbor under $C$, thus, $C(v) = C'(v) = 0$. By a similar argument, for each vertex $v \in V_{nor}$, neighbors of $v$ must all be \ts{nor} vertices, or else, $v$ is adjacent to at least one state-1 vertex (i.e., a $\ts{nand}$ vertex). Moreover, each neighbor of $v$ is not in $V_{nor}$. It follows that the number of state-$0$ neighbors of $v$ is $d(v)$, and $C(v) = C'(v) = 1$. This concludes the correctness for Algorithm~\ref{alg:eqe-nand-nor-memoryless}. As for the time complexity, the for loop from line $2$ to $5$ takes $O(n)$ time, and the for loop from line $6$ to $9$ takes $O(m + n)$ time. Therefore, the overall running time of Algorithm~\ref{alg:eqe-nand-nor-memoryless} is $O(m+n)$. The results for \snsyacg{} and \snsqacg{} follow.
\end{proof}

\begin{algorithm}
\caption{\texttt{EQF\_NAND\_NOR\_SN}($\mc{S'}$)}\label{alg:eqe-nand-nor-memoryless}

\hspace*{0pt} \textbf{Input:} A \snsyds{} $\Bar{\mc{S}} = (G_{\Bar{\mc{S}}}, \mc{F})$, where $\tau_1(v) = 1 \text{ or } d(v), \; \forall v \in V(G_{\Bar{\mc{S}}})$\\
\hspace*{0pt} \textbf{Output:} A fixed point $C$ of $\mc{S}'$
 
\begin{algorithmic}[1]
\State $C \gets $ an initial configuration of all $0$'s

\For{$v \in V(G_{\mc{S}'})$}
    \State \textbf{if} {$\tau_1(v) = 1$} \textbf{then} $C(v) \gets 1$ \Comment{$v$ is a \textsc{nand} vertex}
    
    \State \textbf{else if} {$\tau_1(v) = d(v)$} \textbf{then} $C(v) \gets 0$ \Comment{$v$ is a \textsc{nor} vertex}
\EndFor

\For{$v \in V(G_{\mc{S}'})$}
    \State \textbf{if} $\tau_1(v) = 1$ \textbf{and} $C(w) = 1$ for all neighbors $w \in N(v)$ \textbf{then} $C(v) \gets 0$ \Comment{$V_{nand} = V_{nand} \cup \{v\}$}
    
    \State \textbf{if} $\tau_1(v) = d(v)$ \textbf{and} $C(w) = 0$ for all neighbors $w \in N(v)$ \textbf{then} $C(v) \gets 1$ \Comment{$V_{nor} = V_{nor} \cup \{v\}$}
\EndFor

\State \Return{$C$}
\end{algorithmic}
\end{algorithm}

\subsection*{The underlying graph is a DAG}
Under a directed graph, at each time step, an agent in the \withmemory{} mode considers its own state and the state of \textit{in-neighbors}. Similarly, each agent in the \withoutmemory{} mode only considers the states of its \textit{in-neighbors}.

\begin{mybox2}
\begin{customthm}{{4.9}}
For both the \se{} anti-coordination game and the \sn{} anti-coordination game, an NE can be found in $O(m + n)$ time if the underlying graph is a DAG.
\end{customthm}
\end{mybox2}

\begin{algorithm}
\caption{\texttt{EQF\_DAG\_SE}($\mc{S}$)}\label{alg:eqe-dag-memoryful}

\hspace*{0pt} \textbf{Input:} A \sesyds{} $\mathcal{S} = (\sydsG{}, \mathcal{F})$, where $\sydsG{}$ is a DAG\\
\hspace*{0pt} \textbf{Output:} A fixed point $C$ of $\mc{S}$ (if exists)
 
\begin{algorithmic}[1]
\State $C \gets $ an initial configuration of all $0$'s

\State $V' \gets$ the set of vertices with $0$ indegree in $\sydsG{}$

\For{$v \in V'$}
    \If{$\tau_1(v) = 1$}
     \Return{\textsc{Null}} \Comment{The system has no fixed points}
     \ElsIf{$\tau_1(v) = 0$} 
     $C(v) \gets 1$
     \ElsIf{$\tau_1(v) = 2$} 
     \State $C(v) \gets 0$
     \For{$w \in N(v)$} 
     \State \textbf{if} $\tau_1(w) \neq 0$ \textbf{then} $\tau_1(w) \gets \tau_1(w) - 1$
     \EndFor
    \EndIf
    \State $V' \gets V' \setminus \{v\}$
    \State $V'' \gets$ the set of new vertices with $0$ in-degree
    \State $V' \gets V' \cup V''$
\EndFor

\State \Return{$C$}
\end{algorithmic}
\end{algorithm}

\begin{proof}
We first present the result for \sesyacg{} modeled by a \sesyds{} $\cals{} = (\sydsG{}, \mc{F}')$. Let $V'$ be the set of vertices with $0$ indegree in $\sydsG{}$. Note that $V' \neq \emptyset$, or else, $\sydsG{}$ contains directed cycles. We first remark that $\mc{S}$ has no fixed points if there exists a vertex $v \in V'$ that is not a constant vertex (i.e., $\tau_1(v) = 1, \; \exists v \in V'$). For contrapositive, suppose $\tau_1(v) = 1$ for some $v \in V'$. Given any configuration $C$ where $C(v) = 1$, it follows that $C'(v) = 0$. Similarly, if $C(v) = 0$, then $C'(v) = 1$ for such a vertex $v$. Thus, any configuration $C$ of $\mc{S}$ cannot be a fixed point. 

\par Suppose all vertices in $V'$ are constant vertices. Our algorithm constructs a potential fixed point $C$ as follows. For each vertex $v \in V'$, if $\tau_1(v) = 0$ (i.e., $v$ is a constant-1 vertex), we set $C(v) = 1$, and remove $v$ from $\sydsG{}$. On the other hand, if $\tau_1(v) = 2$ (i.e., $v$ is a constant-0 vertex), we set $C(v) = 0$, then decrease the threshold values of all $v$'s out-neighbors by 1 and remove $v$ from $\sydsG{}$. Note that after removing $v$, we might set some other vertices in $V(\sydsG{}) \setminus V'$ to have $0$ in-degree. Subsequently, we add these new $0$ in-degree vertices to $V'$. The pseudocode is shown in Algorithm~\ref{alg:eqe-dag-memoryful}.

\par If at any iteration of the algorithm, we found a vertex in $V'$ to have threshold $1$ (i.e., a non-constant vertex), the algorithm terminates and concludes that $\mc{S}$ has no fixed points. On the other hand, if all vertices in $V'$ are constant vertices, it follows that the algorithm uniquely determines a fixed point $C$ of $\mc{S}$. As for the running time, we may consider the algorithm as a breadth-first search process that takes $O(m+n)$ time. This concludes the proof \sesyacg{} (and thus \sesqacg{}).

\par As for the \sn{} anti-coordination games, note that a 0-indegree vertex $v$ is always a constant vertex, irrespective of $\tau_1(v)$. Thus, a \sn{} anti-coordination game (either \snsyacg{} or \snsqacg{}) has a unique equilibrium which is determined by the threshold of each vertex. This concludes the proof.
\end{proof}

\subsection*{The underlying graph has no even cycles}

\noindent
We first introduce three characterizations of vertices.

\begin{mybox2}
\begin{definition}[\textbf{Terminal vertices}]
Let $G$ be an undirected graph with no even cycles. A vertex $v$ is a \textbf{terminal vertex} if $v$ is not in any cycles, and $v$ is on the path between two cycles.
\end{definition}
\end{mybox2}

\begin{mybox2}
\begin{definition} [\textbf{Gate vertices}]
Let $G$ be an undirected graph with no even cycles. A vertex $v$ is a \textbf{gate vertex} if $v$ is on at least one cycle and either $(i)$ adjacent to a terminal vertex or $(ii)$ adjacent to another vertex on another cycle, or $(iii)$ on at least two cycles.
\end{definition}
\end{mybox2}

\noindent
Intuitively, \textit{gate vertices} for a cycle $\mc{C}$ act as ``entrances'' on $\mc{C}$, such that a transversal from any other cycles to $\mc{C}$ must reach one of the gate vertices.

\begin{mybox2}
\begin{definition}[\textbf{Tree vertices}]
Let $G$ be an undirected graph with no even cycles. A vertex $v$ is a \textbf{tree vertex} if $v$ is not on any cycles and $v$ is not a terminal vertex.
\end{definition}
\end{mybox2}

\begin{figure}[!h]
    \centering
    \includegraphics[width=0.7\textwidth]{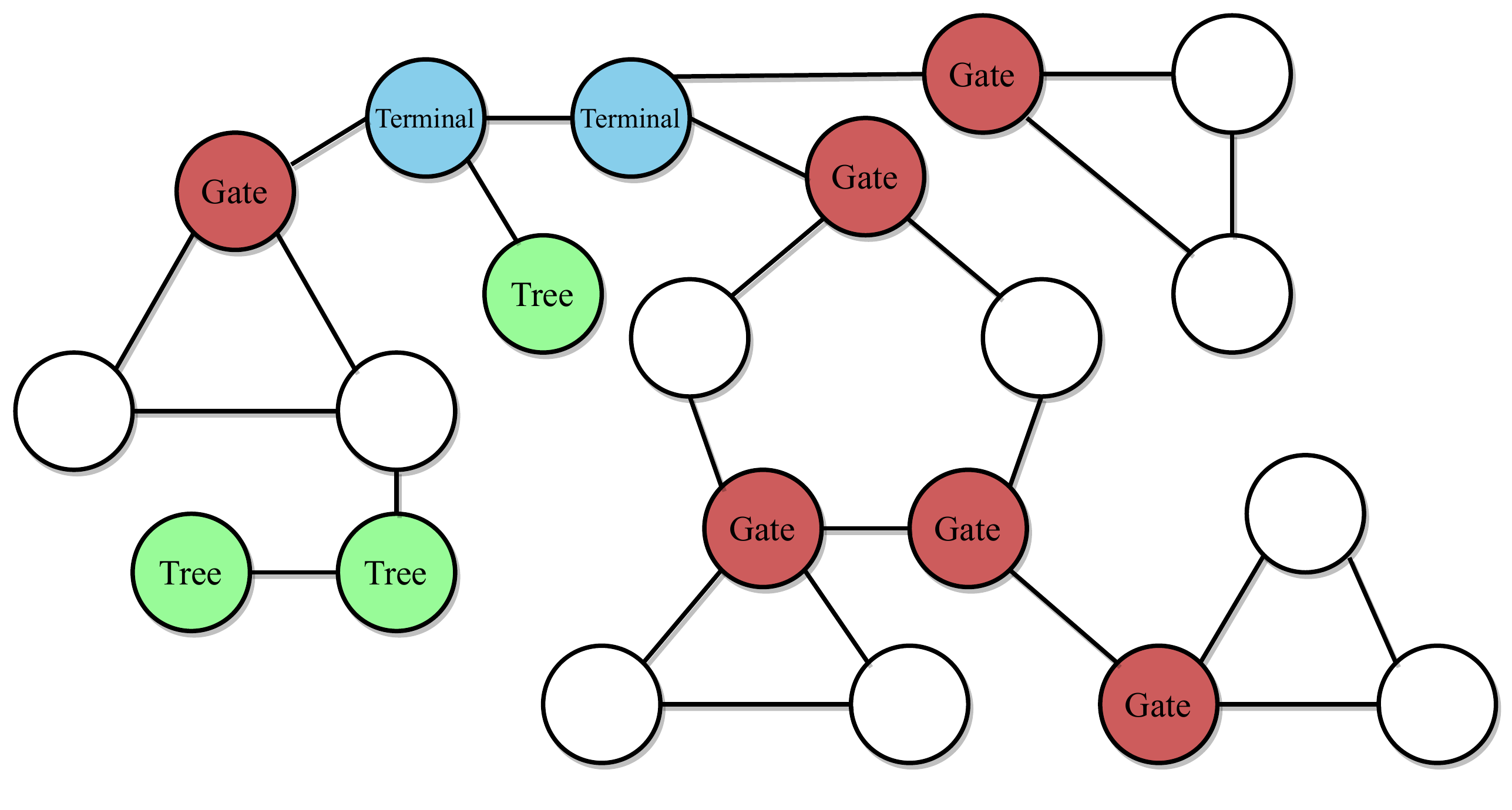}
    \caption{An example graph $G$ where blue vertices are \textit{terminals}, green vertices are \textit{gates}, and red vertices are \textit{tree} vertices.}  
    \label{fig:no-even-cycle}
\end{figure}

\begin{mybox2}
\begin{customlemma}{{4.10}}\label{lemma:odd-cycle-edge-disjoint}
Given an undirected graph $G$ with no even cycles, all cycles in $G$ are edge-disjoint.
\end{customlemma}
\end{mybox2}

\begin{proof}
For contrapositive, suppose there exists two different odd cycles, denoted by $\mc{C}$ and $\mc{C}'$, such that they share a common path $\mc{L} = (v_1, ..., v_l)$ of length $l \geq 1$. Observe that the set of vertices $V(\mc{C}) \cup V(\mc{C}') \setminus V(\mc{L})$ form another cycle, denoted by $\mc{C}''$. Let $2k + 1$ and $2k' + 1$ be the length of $\mc{C}$ and $\mc{C}'$, respectively, $k, k' \geq 1$. It follows that the length of $\mc{C}''$ is $2k + 1 + 2k' + 1 - 2l = 2(k + k' + 1 - l)$ which is even. Proved by contraposition.
\end{proof}

\begin{mybox2}
\begin{customlemma}{{4.11}}\label{lemma:one-gate-vertex}
Let $G$ be an undirected graph with no even cycles, then there must exist a cycle $\mc{C}$ with at most one gate vertex.
\end{customlemma}
\end{mybox2}

\begin{proof}
Given any cycle $\mc{C}$, let $v$ be a gate vertex of $\mc{C}$. By the definition of gate vertices, remark that there exists at least one path $\mc{P}$ from $v$ to another cycle\footnote{If $v$ is on more than one cycle, then the path consists of only vertex $v$ itself.}, where $\mc{P}$ intersects with $\mc{C}$ only on $v$. For contradiction, suppose all cycles have at least two gate vertices. Let $\mc{C}_1$ be any cycle in $G$, and denoted by $v_1$ a gate vertex of $\mc{C}_1$. We label all vertices in $C_1$ as \textit{visited}, and consider a depth-first search (DFS) process from $v_1$ that traverses unvisited vertices until another gate vertex is found (i.e., a new cycle is reached). 

During the DFS process, we label traversed vertices as \textit{visited}. Let $v_2$ be the gate vertex that the DFS (from $v_1$) encounters, and let $\mc{C}_2$ be a cycle that contains $v_2$. We label vertices in $\mc{C}_2$ also as visited. Since all cycles have at least $2$ gate vertices, let $v_2'$ be another gate vertex of $\mc{C}_2$. Remark that there must \textbf{not} exists a path that consists of only \textit{unvisited} vertices (excepts the two endpoints) from $v_2'$ to any visited vertices, or else, there exists a cycle that share common edges with $\mc{C}_2$ which contradicts $G$ being even-cycle free (Lemma~\ref{lemma:odd-cycle-edge-disjoint}). It follows that if we continue the DFS process from $v_2'$ while traversing unvisited vertices, it must reach a new gate vertex $v_3$ on a cycle $\mc{C}_3$. Let $v_3'$ be another gate vertex of $\mc{C}_3$. Similarly, there must \textbf{not} exists a path that consists of only \textit{unvisited} vertices (except the two endpoints) from $v_3'$ to any visited vertices. Overall, given a newly visited gate vertex $v_k$ on a cycle $C_k$, $k \geq 2$. Let $v_k'$ be another gate vertex of $C_k$. By induction, there exists a path with only unvisited vertices (except $v_k'$ itself who is visited) from $v_k'$ to a cycle that is different from $C_i$, $i = 1, ..., k-1$. However, this contradicts $G$ being finite (and thus the number of cycles is finite). The Lemma follows.
\end{proof}

\noindent
Now consider a \sesyds{} $\mc{S} = (\sydsG{}, \mc{F})$ that models a \sesyacg{}.

\begin{mybox2}
\begin{customlemma}{{4.12}}\label{lemma:tree-vertex}
Let $\mc{S} = (\sydsG{}, \mc{F})$ be a \sesyds{} where the underlying graph has no even cycles. A tree vertex $v \in V(\sydsG{})$ has the same state over any fixed point $C$.
\end{customlemma}
\end{mybox2}

\begin{proof}
We first argue that if a vertex $v$ has degree $1$, then $v$ has the same state under any fixed point $C$. In particular, note that the threshold $\tau_1(v)$ could be either $0$, $1$, $2$ or $3$. If $\tau_1(v) = 0$ or $\tau_1(v) = 3$, then $v$ is a constant vertex whose state is uniquely determined by $\tau_1(v)$. On the other hand, by Observation~\ref{obs:threshold_1} and~\ref{obs:threshold_degree}, we know that $C(v) = 1$ if $\tau_1(v) = 1$ and $C(v) = 0$ if $\tau_1(v) = 2$. 

\par Observe that if there exists at least one tree vertex in $\sydsG{}$, then at least one of the tree vertices has degree $1$. Let $v$ be any tree vertex with degree $1$. Based on the claim above, we know that the state of $v$ under any fixed point $C$ is predetermined. Subsequently, we can effectively remove $v$ from the graph and update the threshold values of $v$'s neighbor accordingly (i.e., decrease the neighbor's threshold by one if $C(v) = 0$). By recursion, it follows that the state of any tree vertex $v$ is the same over any fixed point of $\mc{S}$.
\end{proof}

\begin{mybox2}
\begin{customlemma}{{4.13}}
Let $\mc{S} = (\sydsG{}, \mc{F})$ be a \sesyds{} where the underlying graph has no even cycles. Let $\mc{C}$ be a cycle in $\sydsG{}$ with at most one gate vertex. Let $u \in \mc{C}$ be a non-gate vertex on $\mc{C}$, then the state of $u$ is the same under any fixed point.
\end{customlemma}
\end{mybox2}

\begin{proof}
By Lemma~\ref{lemma:one-gate-vertex}, we know that such a cycle $\mc{C}$ with at most one gate vertex must exist. Given a non-gate vertex $u$ on $\mc{C}$, observe that any of $u$'s neighbors that are not on $\mc{C}$ must be tree vertices. Furthermore, by Lemma~\ref{lemma:tree-vertex}, the states of all tree vertices are predetermined under a fixed point. Thus, we can consider the graph where all the tree vertices are removed, and the threshold values of their non-tree neighbors are updated accordingly. Subsequently, all non-gate vertices on $\mc{C}$ have degree $2$.

\begin{customclaim}{{4.13.1}}\label{claim:non-gate-tau-2}
If $\mc{S}$ has a fixed point, then at least one non-gate vertex on $\mc{C}$ does not have threshold $2$.
\end{customclaim}

For contradiction, suppose all the non-gate vertices on $\mc{C}$ have thresholds $2$. Since $\mc{C}$ is of odd length, it cannot be 2-colored. Thus, under any configuration $C$, there must exist an edge $(u, v)$ on $\mc{C}$ where the state of $u$ and $v$ are the same. Furthermore, at least one of them is a non-gate vertex with threshold $2$. W.o.l.g., let $u$ be such a vertex. Since $u$ has degree 2, if $C(u) = C(v) = 1$, then the number of state-$0$ vertices in the closed neighborhood of $u$ is at most $1$, which is less than $\tau_1(u)$. Thus, $C(u) = 1 \neq C'(u) = 0$ where $C'$ is the successor of $C$. Similarly, if $C(u) = C(v) = 0$, we have $C(u) = 0 \neq C'(u) = 1$. Overall, it follows that $\mc{S}$ does not have a fixed point if all the non-gate vertices on $\mc{C}$ have thresholds $2$.

\par Claim~\ref{claim:non-gate-tau-2} implies that if $\mc{S}$ has a fixed point, then at least one non-gate vertex $u$ on $\mc{C}$ has a threshold not equal to $2$. We now argue that the state of $u$ remains the same under any fixed point $C$ of $\mc{S}$. In particular, if $\tau_1(u) = 0$ or $\tau_1(u) = 4$, then $u$ is a constant vertex whose state is predetermined. On the other hand, if $\tau_1(u) = 1$, then $C(u) = 1$, and if $\tau_1(u) = 3 = d(v) + 1$, then $C(u) = 0$. It follows that the state of $u$ is the same under any fixed point and is predetermined by $\tau_1(v)$. Thus, we may remove $u$ from the graph and update the thresholds of $u$'s neighbors accordingly. Remark that after removing $u$, all non-gate vertices on $\mc{C}$ become tree vertices. Subsequently, by Lemma~\ref{lemma:tree-vertex}, the state of all non-gate vertices on $\mc{C}$ are predetermined, and are the same over any fixed point of $\mc{C}$.
\end{proof}

\begin{mybox2}
\begin{customthm}{{4.14}}
For a \se{} anti-coordination game, an NE can be found in $O(m + n)$ time if the underlying graph has no even cycles.
\end{customthm}
\end{mybox2}

\begin{proof}
Let $\cals{} = (\sydsG{}, \mc{F})$ be a \sesyds{}{} that models a \sesyacg{}. Suppose $\mc{S}$ has a fixed point, denoted by $C$. By Lemma~\ref{lemma:tree-vertex}, we can determine the state of all the tree vertices under $C$ based on their threshold values, and remove all the tree vertices from $\sydsG{}$. Let $\mc{C}_1$ be a cycle that consists of only one gate vertex. By Lemma~\ref{lemma:one-gate-vertex}, we can also determine the states of all non-gate vertices on $\mc{C}_1$ and effectively eliminate the cycle $\mc{C}_1$. Let $\sydsG{}_1$ be the resulting graph which is still even-cycle free. If $\sydsG{}_1$ contains other cycles, then there must exists another cycle that consists of only one gate vertex. By recursively determining the state of non-gate vertices on single-gate cycles and the states of tree vertices, it follows that we can determine the states of all vertices under $C$ in time $O(m + n)$. Since a fixed point of $\mc{S}$ corresponds to an NE of the underlying \sesyacg{}, we conclude that an NE can be found in $O(m + n)$ time for a \sesyacg{} (and thus for \sesqacg{}) when the underlying graph has no even cycles.
\end{proof} 

\noindent
With the same argument, we also establish the same result for \sn{} anti-coordination games.

\begin{mybox2}
\begin{customthm}{{4.15}}
For a \sn{} anti-coordination game, an NE can be found in $O(m + n)$ time if the underlying graph has no even cycles.
\end{customthm}
\end{mybox2}

\subsection*{The underlying graph is complete}

\begin{mybox2}
\begin{customthm}{{4.16}}
For a \se{} anti-coordination game, an NE can be found in $O(n)$ time if the underlying graph is a complete graph.
\end{customthm}
\end{mybox2}

\begin{proof}
Let $\cals{} = (\sydsG{}, \mc{F})$ be a \sesyds{} that models a \sesyacg{}. We partition the set of vertices based on their thresholds. Specifically, let $k$ be the number of distinct thresholds $\tau_1$. Let $\mc{P} = \{V_1, ..., V_k\}$ be a partition of the vertex set $V(\sydsG{})$, such that $\tau_1(u) = \tau_1(v), \; \forall u, v \in V_i, \; i = 1, ..., k$. Furthermore, $\tau_1(u) < \tau_1(v), \; \forall u \in V_i, v \in V_j, \; 1 \leq i < j \leq k$. Remark that since $\sydsG{}$ is a complete graph, the closed neighborhoods of all vertices are the same, which is $V(\sydsG{})$. Thus, given any fixed point $C$ of $\mc{S}$, if $C(v) = 1$ for some $v \in V_j$, then $C(u) = 1, \forall u \in V_i, i \leq j$.

\par Our algorithm consists of at most $k - 1$ iterations, where in each iteration $j$, $1 \leq j \leq k - 1$, we construct a configuration $C$ such that $C(v) = 1, \forall v \in V_i, i = 1, ..., j$, and $C(v) = 0$ otherwise. After each iteration of the algorithm, we check if the resulting $C$ is a fixed point, and return $C$ if so. On the other hand, if all the resulting $k - 1$ configurations are not fixed points, we conclude that $\mc{S}$ does not have a fixed point. The pseudocode is shown in Algorithm~\ref{alg:eqe-complete-memoryful}.

\noindent
We establish the correctness of the algorithm. First, we claim that

\begin{customclaim}{{4.16.1}}\label{claim:vp-vpp}
The system $\mc{S}$ has a fixed point if and only if there exists a bipartition $\{V', V''\}$ of $V(\sydsG{})$, such that $\max \{\tau_1(v) : v \in V'\} \leq |V''| < \min \{\tau_1(v) : v \in V''\}$. Further, such a fixed point $C$ has vertices in $V'$ in state $1$, and vertices in $V''$ in state $0$.
\end{customclaim}

Suppose $\mc{S}$ has a fixed point $C$. Let $C'$ be the successor of $C$. Let $V' = \{v : C(v) = 1, v \in V(\sydsG{})\}$ and $V'' = \{v : C(v) = 0, v \in V(\sydsG{})\}$ be a bipartition of the vertex set based on the state of vertices. We argue that $\max \{\tau_1(v) : v \in V'\} < \min \{\tau_1(v) : v \in V''\}$. For contrapositive, assume that there exists a vertex $v \in V'$ such that $\tau_1(v) \geq \tau_1(w)$ for some $w \in V''$. Since $C$ is a fixed point and $C(v) = 1$, we have $|V''| \geq \tau_1(v)$ (i.e., the number of state-0 vertices in $C$ is at least $\tau_1(v)$). However, it follows that $|V''| \geq \tau_1(v) \geq \tau_1(w)$, thus, $C(w) = 0 \neq C'(w) = 1$ and $C$ is not a fixed point. Overall, we have $\max \{\tau_1(v) : v \in V'\} \leq |V''| < \min \{\tau_1(v) : v \in V''\}$. 

\par As for the necessity of the claim, note that if such a bipartition $\{V', V''\}$ exists, we can construct a fixed point $C$ by assigning $C(v) = 1, \; \forall v \in V'$ and $C(v) = 0 \; \forall v \in V''$. It follows that $|V''| \geq \max \{\tau_1(v) : v \in V'\} \geq \tau_1(v), \; \forall v \in V'$. Thus, $C(v) = C'(v) = 1$, $\forall v \in V'$. Similarly, we have $|V''| \leq \min \{\tau_1(v) : v \in V''\} \leq \tau_1(v), \; \forall v \in V''$ and $C(v) = C'(v) = 0$, $\forall v \in V''$. This concludes Claim~\ref{claim:vp-vpp}. 

We now argue that the \textit{for loop} of Algorithm~\ref{alg:eqe-complete-memoryful} from line $5$ to $12$ essentially discovers if such a bipartition $\{V', V''\}$ exists. In particular, at the $j^{\text{th}}$ iteration, $1 \leq j \leq k-1$, denoted by $a_0$ the number of state-$0$ vertices in $C$ (after the updates from line 6 to 8). Let $\tau_1(V_j)$ be the threshold value of vertices in $V_j \in \mc{P}$. Observe that the algorithm effectively construct a configuration $C$ by setting vertices in $V' = \bigcup_{i = 1}^j V_i$ to state $1$, and vertices in $V'' = \bigcup_{i = j+1}^k V_i$ to state $0$ .

\begin{customclaim}{{4.16.2}}
The inequality $\tau_1(V_j) \leq a_0 < \tau_1(V_{j+1})$ (i.e., the condition at line 10) is satisfied if and only if $\max \{\tau_1(v) : v \in V'\} \leq |V''| < \min \{\tau_1(v) : v \in V''\}$. 
\end{customclaim}

Observe that $a_0 = |V''|$. Suppose $\tau_1(V_j) \leq a_0 < \tau_1(V_{j+1})$. Since $\tau_1(V_j) \geq \tau_1(v), \; \forall v \in V' =  \bigcup_{i = 1}^j V_i$, it follows that $a_0 = |V''| \geq \max \{\tau_1(v) : v \in V'\}$. Similarly, given that $\tau_1(V_{j+1}) \leq \tau_1(v), \; \forall v \in V'' =  \bigcup_{j = i+1}^k V_j$, we have $|V''| \leq \min\{\tau_1(v) : v \in V''\}$. For the other direction, by an analogous argument, it is easy to see that if $\max \{\tau_1(v) : v \in V'\} \leq |V''| < \min \{\tau_1(v) : v \in V''\}$, then $\tau_1(V_i) \leq a_0 < \tau_1(V_{i+1})$. The correctness of the algorithm immediately follows.

Overall, the partition $\mc{P}$ can be constructed in $O(n)$ time via the bucket sort. Furthermore, the \textit{for loop} at line $7$ is called at most $O(n)$ time, and the operations from line $9$ to line $12$ take constant time. Therefore, the overall running time is $O(n)$. Since a fixed point of $\mc{S}$ corresponds to an NE of the underlying game, we conclude
that an NE can be found in $O(n)$ time for a \sesyacg{} (and therefore for \sesqacg{}) when the underlying graph is a complete graph.
\end{proof}

\begin{algorithm}
\caption{\texttt{EQF\_Complete\_SE}($\mc{S}$)}\label{alg:eqe-complete-memoryful}

\hspace*{0pt} \textbf{Input:} A \sesyds{}/\textsf{SDS} $\mathcal{S} = (G_{\mathcal{S}}, \mathcal{F})$, where $\sydsG{}$ is a complete graph\\
\hspace*{0pt} \textbf{Output:} A fixed point $C$ of $\mc{S}$
 
\begin{algorithmic}[1]
\State $C \gets $ an initial configuration of all $0$'s

\State $\mc{P} \gets \{V_1, ..., V_k\}$ be a partition of $V(\sydsG{})$ such that $\tau_u < \tau_v$ iff $u \in V_i, v \in V_j, 1 \leq i < j \leq k$

\State $\tau_1(V_i) \gets$ the threshold value of vertices in $V_i \in \mc{P}, i = 1, ..., k$

\State $a_0 \gets |V(\sydsG{})|$ \Comment{The number of state-$0$ vertices in $C$}

\For{$j = 1$ \textbf{to} $k-1$}
\For{$v \in V_j$}
    \State $C(v) \gets 1$ 
\EndFor

\State $a_0 \gets a_0 - |V_j|$

\If{$\tau_1(V_j) \leq a_0 < \tau_1(V_{j+1})$} \Comment{Determine if $C$ is a fixed point}

\Return{C}
\EndIf
\EndFor

\State \Return{\ts{Null}} \Comment{The system has no fixed point}
\end{algorithmic}
\end{algorithm}

\noindent
With the same argument, we can show that the above result for complete graphs carries over to \sn{} anti-coordination games.

\begin{mybox2}
\begin{customthm}{{4.17}}
For a \sn{} anti-coordination game, an NE can be found in $O(n)$ time if the underlying graph is a complete graph.
\end{customthm}
\end{mybox2}

\section*{Additional Material for Section 5}
In this section, we present the detailed proofs of the convergence result for \textit{synchronous} anti-coordination games, given in section 5 of the main manuscript. Based on the connection between the games and dynamical systems, all proofs are given in the context of \textit{synchronous} dynamical systems. We first present the convergence result for \snsyds{} (modeling \snsyacg{}). Then with simple modifications of the proof, the result for \sesyds{} (modeling \sesyacg{}) follows. Let $\cals{} = (\sydsG{}, \mc{F})$ be a \snsyds{} that models a \snsyacg{}. Recall that for a vertex $v$, $\tau_0(v)$ is the minimum number of state-$1$ neighbors of $v$ for $f_v$ to equal $0$. W.l.o.g., we assume there are no constant vertices. That is, $1 \leq \tau_1(u), \tau_0(u) \leq d(u)$ for all $u \in V(\sydsG{})$.

\subsection*{The potential functions and bounds} Let $\Ct{}$ be an arbitrary configuration of $\mc{S}$ whose successor is $\Ct{'}$. We now define the potentials of vertices, edges, and the system under $\Ct{}$. For each vertex $u \in V(\sydsG{})$, let $\Tilde{\tau}_0(u) = \tau_0(u) - 1/2$. 

\noindent
\underline{Vertex potential}. Given a vertex $u \in V(\sydsG{})$, the potential of $u$ under $\Ct{}$, is defined as follows: 
$$
\mc{P}(\Ct{}, u) = \Ct{}(u) \cdot \Tilde{\tau}_0(u) + \Ct{'}(u) \cdot \Tilde{\tau}_0(u)
$$

\noindent
\underline{Edge potentials}. Given an edge $e = (u, v) \in E(\sydsG{})$, the potential of $e$ under $\Ct{}$ is defined as follows:
$$
    \mc{P}(\Ct{}, e) = \Ct{}(u) \cdot \Ct{'}(v) + \Ct{}(v) \cdot \Ct{'}(u)
$$

\noindent
\underline{The configuration potential.} The potential of the system $\mc{S}$ under $\Ct{}$ is defined as the subtraction of the sum of vertex potentials from the sum of edge potentials:
$$
    \mc{P}(\Ct{}, \mc{S}) = \sum_{e \in E(\sydsG{})} \mc{P}(\Ct{}, e) - \sum_{u \in V(\sydsG{})} \mc{P}(\Ct{}, u)
$$

\vspace{-0.5cm}
\paragraph{A lower bound on the potential.} We first present a lower bound on the configuration potential under $\Ct{}$.

\begin{mybox2}
\begin{customlemma}{{5.1}}
The configuration potential satisfies 
$$\mc{P}(\Ct{}, \mc{S}) \geq -4m + n$$
\end{customlemma}
\end{mybox2}

\begin{proof}
First observe that the sum of edge potentials is non-negative. Further, the sum of potentials of vertices is upper bounded by
\begin{align}
    \sum_{u \in V(\sydsG{})} \mc{P}(\Ct{}, u) &= \sum_{u \in V(\sydsG{})} \Ct{}(u) \cdot \Tilde{\tau}_0(u) + \Ct{'}(u) \cdot \Tilde{\tau}_0(u) \nonumber \\
    &\leq 2 \sum_{u \in V(\sydsG{})} \Tilde{\tau}_0(u)\\
    &\leq 2 \sum_{u \in V(\sydsG{})} (d(u) - \frac{1}{2}) \nonumber\\
    &= 4m - n \nonumber
\end{align}
The lower bound of the configuration potential immediately follows.
\end{proof}

\vspace{-0.5cm}
\paragraph{An upper bound on the potential.} We now present an upper bound on the configuration potentials under $\Ct{}$.

\begin{mybox2}
\begin{customlemma}{{5.2}}
The configuration potential $\mc{P}(\Ct{}, \mc{S})$ is upper bounded by $0$.
\end{customlemma}
\end{mybox2}

\begin{proof}
Let $E_u = \{(u, v) : (u, v) \in E(\sydsG{})\}$ be the set of edges incident to a vertex $u \in V(\sydsG{})$. We can restate the sum of edge potentials as follows
\begin{align}
    \sum_{e \in E(\sydsG{})} \mc{P}(\Ct{}, e) &= \sum_{(u,v) \in E(\sydsG{})} \left(\Ct{}(u) \cdot \Ct{'}(v) + \Ct{}(v) \cdot \Ct{'}(u)\right) \\
    &= \sum_{u \in V(\sydsG{})} \left(\Ct{'}(u) \cdot \sum_{(u, v) \in E_u} \Ct{}(v)\right) \nonumber
\end{align}
We can further expend the configuration potential into the form
\begin{align}
    \mc{P}(\Ct{}, \mc{S}) = \sum_{u \in V(\sydsG{})} \left(\left(\Ct{'}(u) \sum_{(u, v) \in E_u} \Ct{}(v)\right) - \Ct{}(u) \cdot \Tilde{\tau}_0(u) - \Ct{'}(u) \cdot \Tilde{\tau}_0(u)\right)
\end{align}
Note that $\sum_{(u, v) \in E_u} \Ct{}(v)$ is exactly the number of state-$1$ neighbors of $u$ under $\Ct{}$. If $\Ct{'}(u) = 0$, then 
\begin{align}
    \left(\Ct{'}(u) \sum_{e = (u, v) \in E_u} \Ct{}(v)\right) - \Ct{}(u) \cdot \Tilde{\tau}_0(u) - \Ct{'}(u) \cdot \Tilde{\tau}_0(u) &= - \Ct{}(u) \cdot \Tilde{\tau}_0(u) \leq 0
\end{align}

Conversely, if $\Ct{'}(u) = 1$, then by the inverted-threshold dynamics, the number of state-$1$ neighbor of $u$ under $\Ct{}$ is less than $\tau_0(u)$. Subsequently, we have
\begin{equation}
    \sum_{(u, v) \in E_u} \Ct{}(v) \leq \tau_0(u) - 1 < \Tilde{\tau}_0(u) 
\end{equation}
and 
\begin{align}
    &\left(\Ct{'}(u) \sum_{(u, v) \in E_u} \Ct{}(v)\right) - \Ct{'}(u) \cdot \Tilde{\tau}_0(u) - \Ct{}(u) \cdot \Tilde{\tau}_0(u) \\
     &= \left(\sum_{(u, v) \in E_u} \Ct{}(v)\right) - \Tilde{\tau}_0(u) - \Ct{}(u) \cdot \Tilde{\tau}_0(u)\nonumber \\
     &\leq 0 \nonumber
\end{align}
It follows that the overall configuration potential satisfies $\mc{P}(\Ct{}, \mc{S}) \leq 0$.
\end{proof}

\noindent
Overall, we have established that the gap in the configuration potential value between two configurations is at most $4m - n$.

\subsection*{Decrease of the configuration potential before convergence}
In this section, we show that from a configuration $\Ct{}$, the configuration potential decrease by at least $1$ after every two time-steps, until a fixed point or a 2-cycle is reached.  

\begin{mybox2}
\begin{customlemma}{{5.3}}\label{lemma:convergence}
Let $\Ct{}$ an arbitrary configuration of $\mc{S}$. We have $\mc{P}(\Ct{'}, \mc{S}) = \mc{P}(\Ct{}, \mc{S})$ if and only if $\Ct{} = \Ct{''}$, that is $\Ct{}$ is a fixed point or is on a $2$-cycle $\Ct{} \longleftrightarrow \Ct{'}$. Furthermore, if $\Ct{} \neq \Ct{''}$ (i.e., the dynamic has not converged), then $\mc{P}(\Ct{'}, \mc{S}) - \mc{P}(\Ct{}, \mc{S}) \leq - 1/2$.
\end{customlemma}
\end{mybox2}

\begin{proof}
We defined the change of potential of an edge $e = (u, v) \in E(\sydsG{})$ from $\Ct{}$ to $\Ct{'}$ as 
\begin{equation}
    \Delta(e) = \mc{P}(\Ct{'}, e) - \mc{P}(\Ct{}, e)
\end{equation}
and the change of potential of a vertex $u$ as
\begin{equation}
\Delta(u) = \mc{P}(\Ct{'}, u) - \mc{P}(\Ct{}, u)
\end{equation}

\noindent
Subsequently, the change of the configuration potential from $\Ct{}$ to $\Ct{'}$ is as follows
\begin{align}
    &\mc{P}(\Ct{'}, \mc{S}) - \mc{P}(\Ct{}, \mc{S})\\
    &= \sum_{e \in E(\sydsG{})} \mc{P}(\Ct{'}, e) - \sum_{u \in V(\sydsG{})} \mc{P}(\Ct{'}, u) - \sum_{e \in E(\sydsG{})} \mc{P}(\Ct{}, e) + \sum_{u \in V(\sydsG{})} \mc{P}(\Ct{}, u) \nonumber \\
    &= \sum_{e \in E(\sydsG{})} \Delta(e) - \sum_{u \in V(\sydsG{})} \Delta(u) \nonumber
\end{align}

We now expand $\Delta(e)$ for $e = (u,v)$ and $\Delta(u)$ as follows
\begin{align}
    \Delta(e) &= \left(\Ct{'}(u) \cdot \Ct{''}(v) + \Ct{'}(v) \cdot \Ct{''}(u)\right) - \left(\Ct{}(u) \cdot \Ct{'}(v) + \Ct{'}(u) \cdot \Ct{}(v)\right) \\
    &= \Ct{'}(u) \cdot \Ct{''}(v) + \Ct{'}(v) \cdot \Ct{''}(u) - \Ct{}(u) \cdot \Ct{'}(v) - \Ct{'}(u) \cdot \Ct{}(v) \nonumber \\
    &= \Ct{'}(u) \cdot \left(\Ct{''}(v) - \Ct{}(v)\right) + \Ct{'}(v) \cdot \left(\Ct{''}(u) - \Ct{}(u)\right) \nonumber
\end{align}
\begin{align}
    \Delta(u) &= \left(\Ct{'}(u) \Tilde{\tau}_0(u) + \Ct{''}(u) \Tilde{\tau}_0(u)\right) - \left(\Ct{}(u) \Tilde{\tau}_0(u) + \Ct{'}(u) \Tilde{\tau}_0(u)\right)\\
    &= \left(\Ct{''}(u) - \Ct{}(u)\right) \cdot \Tilde{\tau}_0(u) \nonumber
\end{align}

We argue that the change in configuration potential is $0$ if and only if $\Ct{} = \Ct{''}$. 

\begin{itemize}
    \item[$(\Leftarrow)$] Suppose $\Ct{} = \Ct{''}$, it follows that $\Ct{}(u) = \Ct{''}(u)$, $\forall u \in V(\sydsG{})$. Subsequently, $\Delta(e) = 0$, $\forall e \in E(\sydsG{})$ and $\Delta(u) = 0$, $\forall u \in V(\sydsG{})$. Overall, we conclude that the change of potential $\mc{P}(\Ct{'}, \mc{S}) - \mc{P}(\Ct{}, \mc{S}) = 0$. 
    
    \item[$(\Rightarrow)$] For contrapositive, suppose $\Ct{} \neq \Ct{''}$. We now show that the configuration potential decreases by at least $1/2$. It is easy to see that given a vertex $u$ such that $\Ct{}(u) = \Ct{''}(u)$, we have $\Delta(u) = 0$. Thus, we consider the set of vertices whose states in $\Ct{}$ are different from the states in $\Ct{''}$. Let $V_{0-1} = \{u : \Ct{}(u) = 0, \; \Ct{''}(u) = 1, \; u \in V(\sydsG{})\}$ be the set of vertices whose states are $0$'s under $\Ct{}$ and are $1$'s under $\Ct{''}$. Analogously, denoted by $V_{1-0} = \{u : \Ct{}(u) = 1, \; \Ct{''}(u) = 0, \; u \in V(\sydsG{})\}$ the set of vertices whose states are $1$'s under $\Ct{}$ and are $0$'s under $\Ct{''}$. We remark that since $\Ct{} \neq \Ct{''}$, $|V_{0-1}| + |V_{1-0}| \geq 1$.
    
    \par We now consider the change of potential for a vertex $u$ whose state in $\Ct{}$ is different from the state in $\Ct{''}$, under the following cases.
        \par \underline{Case 1}: The vertex $u \in V_{0-1}$. Subsequently, the change of vertex potential for $u$ is 
        \begin{equation}
            \Delta(u) = \left(\Ct{''}(u) - \Ct{}(u)\right) \cdot \Tilde{\tau}_0(u) = \Tilde{\tau}_0(u)
        \end{equation}
        
        \par \underline{Case 2}: The vertex $u \in V_{1-0}$. It then follows that
        \begin{equation}
            \Delta(u) = \left(\Ct{''}(u) - \Ct{}(u)\right) \cdot \Tilde{\tau}_0(u) = -\Tilde{\tau}_0(u)
        \end{equation}

\par Overall, we have
\begin{equation}\label{eq:change-vex-potential}
    \sum_{u \in V(\sydsG{})} \Delta(u) = \sum_{u \in V_{0-1}} \Tilde{\tau}_0(u) - \sum_{u \in V_{1-0}} \Tilde{\tau}_0(u)
\end{equation}

    As for the change of edge potentials, observe that $\Delta(e) = 0$ for $e = (u, v)$ if $\Ct{}(u) = \Ct{''}(u)$ and $\Ct{}(v) = \Ct{''}(v)$. We now consider $\Delta(e)$, $e = (u, v)$ where either $u$ or $v$ (or both) undergoes state change from $\Ct{}$ to $\Ct{''}$.
    
    \par \underline{Case 1}: The state of one vertex altered, and the state of the other remained the same. W.o.l.g., suppose the vertex $u \in V_{0-1}$, and $\Ct{}(v) = \Ct{''}(v)$. It follows that 
    \begin{align}
        \Delta(e)
        &= \Ct{'}(u) \cdot \left(\Ct{''}(v) - \Ct{}(v)\right) + \Ct{'}(v) \cdot \left(\Ct{''}(u) - \Ct{}(u)\right) \\
        &= \Ct{'}(v) \nonumber
    \end{align}
    On the other hand, if $u \in V_{1-0}$, we then have $\Delta(e) = -\Ct{'}(v)$. Analogously, if the state of $v$ changed and the state of $u$ remained the same, then $\Delta(e) = \Ct{'}(u)$ when $v \in V_{0-1}$, and $\Delta(e) = -\Ct{'}(u)$ when $v \in V_{1-0}$.
    
    \par \underline{Case 2}: The the states of $u, v$ both changed in the same direction from $\Ct{}$ to $\Ct{''}$ (i.e., both from $0$ to $1$ or from $1$ to $0$). Specifically, if $u, v \in V_{0-1}$, we then have
    \begin{align}
        \Delta(e)
        &= \Ct{'}(u) \cdot \left(\Ct{''}(v) - \Ct{}(v)\right) + \Ct{'}(v) \cdot \left(\Ct{''}(u) - \Ct{}(u)\right) \\
        &= \Ct{'}(v) + \Ct{'}(u) \nonumber
    \end{align}
    Conversely, if $u, v \in V_{1-0}$, then $\Delta(e) = -\Ct{'}(v) - \Ct{'}(u)$.
    
    \par \underline{Case 3}: The states of $u, v$ changed in different directions from $\Ct{}$ to $\Ct{''}$ (i.e., one from $0$ to $1$ and the other from $1$ to $0$). W.l.o.g., suppose $u \in V_{0-1}$ and $v \in V_{1-0}$. Subsequently, we have 
    \begin{align}
        \Delta(e)
        = \Ct{'}(v) - \Ct{'}(u)
    \end{align}
    Similarly, if $u \in V_{1-0}$ and $v \in V_{0-1}$, then $\Delta(e) = \Ct{'}(u) - \Ct{'}(v)$
\end{itemize}

\par Overall, observe that for any vertex $u \in V_{0-1}$, the change of edge potential $\Delta(e)$ for each incident edge $e = (u, v)$ has a positive term $\Ct{'}(v)$. Conversely, if $u \in V_{1-0}$, then $\Delta(e)$ has a negative term $-\Ct{'}(v)$. Let $E_u = \{(u, v) : (u, v) \in E(\sydsG{})\}$ be the set of edges incident to $u \in V(\sydsG{})$, we can rewire the sum of the change in edge potentials as 
\begin{align}
    \sum_{e \in E(\sydsG{})} \Delta(e) = \left(\sum_{u \in V_{0-1}} \sum_{(u,v) \in E_u} \Ct{'}(v)\right) - \left(\sum_{u \in V_{1-0}} \sum_{(u,v) \in E_u} \Ct{'}(v)\right)
\end{align}
In combined with Equation~\ref{eq:change-vex-potential}, we characterize the change in the configuration potential from $\Ct{}$ to $\Ct{'}$ as follows
\begin{align}
    &\mc{P}(\Ct{'}, \mc{S}) - \mc{P}(\Ct{}, \mc{S})\\
    &= \sum_{e \in E(\sydsG{})} \Delta(e) - \sum_{u \in V(\sydsG{})} \Delta(u) \nonumber\\
    &= \left(\sum_{u \in V_{0-1}} \sum_{(u,v) \in E_u} \Ct{'}(v) - \sum_{u \in V_{1-0}} \sum_{(u,v) \in E_u} \Ct{'}(v)\right) - \left(\sum_{u \in V_{0-1}} \Tilde{\tau}_0(u) - \sum_{u \in V_{1-0}} \Tilde{\tau}_0(u)\right) \nonumber \\
    &= \left(\sum_{u \in V_{0-1}} \left(\sum_{(u,v) \in E_u} \Ct{'}(v)\right) - \Tilde{\tau}_0(u)\right) + \left(\sum_{u \in V_{1-0}} \Tilde{\tau}_0(u) - \left(\sum_{(u,v) \in E_u} \Ct{'}(v)\right)\right) \nonumber
\end{align}

\par Remark that the term $\sum_{(u,v) \in E_u} \Ct{'}(v)$ is the number of state-$1$ neighbors of a vertex $u$ under $\Ct{'}$. Suppose $u \in V_{0-1}$, that is, $\Ct{''}(u) = 1$. By the dynamics of inverted-threshold model, it follows that the number of state-$1$ neighbors of $u$ under $\Ct{'}$ is less than $\tau_0(u) = \Tilde{\tau}_0(u) + 1/2$, thus,
$$
     \sum_{(u,v) \in E_u} \Ct{'}(v) \leq \tau_0(u) - 1 < \Tilde{\tau}_0(u)
$$
and 
$$
    \left(\sum_{(u,v) \in E_u} \Ct{'}(v)\right) - \Tilde{\tau}_0(u) \leq -\frac{1}{2}
$$

Conversely, if $u \in V_{1-0}$ which implies that $\Ct{''}(u) = 0$. It follows that the number of state-$1$ neighbors of $u$ is at least $\tau_0(u)$. Thus,
$$
     \sum_{(u,v) \in E_u} \Ct{'}(v) \geq \tau_0(u) > \Tilde{\tau}_0(u)
$$
and 
$$
    \Tilde{\tau}_0(u) - \sum_{(u,v) \in E_u} \Ct{'}(v) \leq -\frac{1}{2}
$$
Overall, we have
\begin{align}
    &\mc{P}(\Ct{'}, \mc{S}) - \mc{P}(\Ct{}, \mc{S})\\
    &= \left(\sum_{u \in V_{0-1}} \left(\sum_{(u,v) \in E_u} \Ct{'}(v)\right) - \Tilde{\tau}_0(u)\right) + \left(\sum_{u \in V_{1-0}} \Tilde{\tau}_0(u) - \left(\sum_{(u,v) \in E_u} \Ct{'}(v)\right)\right) \nonumber\\
    &\leq - \sum_{u \in V_{0-1}} \frac{1}{2} - \sum_{u \in V_{1-0}} \frac{1}{2} \nonumber\\
    &= - \frac{|V_{0-1}| + |V_{1-0}|}{2} \nonumber\\
    &\leq -\frac{1}{2} \nonumber
\end{align}
This concludes the proof.
\end{proof}

\noindent
Overall, we have shown that the potential gap between any two configurations is $4m - n$. Furthermore, the overall system configuration decreases by at least $1/2$ after each time step. It follows that for a \snsyds{}, starting from an arbitrary initial configuration, the system dynamic converges in at most $8m - 2n$ time steps, irrespective of the underlying network structure. Subsequently, the convergence time result for \snsyacg{} follows.

\begin{mybox2}
\begin{customthm}{{5.4}}
For \snsyacg{}, starting from any initial action profile, the best-response dynamic converges to a Nash equilibrium or a 2-cycle in $O(m)$ time steps.
\end{customthm}
\end{mybox2}

\noindent
With simple modifications of Lemma~\ref{lemma:convergence} (i.e., consider $u$ as a neighbor of itself), we can also show that the same convergence time can be extended to \sesyacg{}.

\begin{mybox2}
\begin{customthm}{{5.5}}
For a \sesyacg{}, starting from any initial action profile, the best-response dynamic converges to a Nash equilibrium or a 2-cycle in $O(m)$ time steps.
\end{customthm}
\end{mybox2}

\begin{mybox2}
\begin{customcoro}{{5.6}}
For both \sesyacg{} and \snsyacg{}, starting from any initial action profile, the best-response dynamic converges to a Nash equilibrium or a 2-cycle in $O(n)$ time steps if the graph is degree bounded.
\end{customcoro}
\end{mybox2}

\end{document}